\documentclass[a4paper,fleqn,usenatbib]{mnras}
\usepackage{newtxtext,newtxmath}
\usepackage[T1]{fontenc}
\usepackage{ae,aecompl}
\usepackage{graphicx}	
\usepackage{amsmath}	
\usepackage{verbatim}
\usepackage{units}
\usepackage{subfigure}
\usepackage{pdflscape}
\usepackage{svg}
\usepackage{bm}
\usepackage{stfloats}
\usepackage{float}
\usepackage{multicol}
\usepackage{hyperref}
\usepackage{multirow}
\usepackage{wasysym}
\usepackage{longtable}

\usepackage{xcolor}
\bibpunct{(}{)}{;}{a}{}{,}

\renewcommand{\arraystretch}{1.5}

\title[Exoplanet Interiors]{Exoplanet Interior Retrievals: core masses and metallicities from atmospheric abundances}

\author[S. Bloot et al.]{S.~Bloot$^{1,2,3}$\thanks{E-mail: bloot@astron.nl},
Y.~Miguel$^{1,4}$, M.~Bazot$^{5,6}$, and S.~Howard$^{7}$
\\
$^{1}$Leiden Observatory, Leiden University, PO\,Box 9513, 2300 RA, Leiden, The Netherlands\\
$^{2}$ASTRON, Netherlands Institute for Radio Astronomy, Oude Hoogeveensedijk 4, 7991 PD, Dwingeloo, The Netherlands\\
$^{3}$Kapteyn Astronomical Institute, University of Groningen, P.O. Box 800, 9700 AV, Groningen, The Netherlands\\
$^{4}$SRON Netherlands Institute for Space Research, Niels Bohrweg 4, 2333 CA, Leiden, the Netherlands\\
$^{5}$ Heidelberg Institute for Theoretical Studies (HITS gGmbH), Schloss-Wolfsbrunnenweg 35,69118 Heidelberg, Germany\\
$^6$ CITIES, NYUAD Institute, New York University Abu Dhabi, PO Box 129188, Abu Dhabi, United Arab Emirates\\
$^7$ Université Côte d Azur, OCA, Lagrange CNRS, 06304 Nice, France
}
\date{Accepted XXX. Received YYY; in original form ZZZ}

\pubyear{2023}

\begin{document}
\label{firstpage}
\pagerange{\pageref{firstpage}--\pageref{lastpage}}
\maketitle

\begin{abstract}
The mass and distribution of metals in the interiors of exoplanets are essential for constraining their formation and evolution processes. Nevertheless, with only masses and radii measured, the determination of exoplanet interior structures is degenerate, and so far simplified assumptions have mostly been used to derive planetary metallicities. In this work, we present a method based on a state-of-the-art interior code, recently used for Jupiter, and a Bayesian framework, to explore the possibility of retrieving the interior structure of exoplanets. We use masses, radii, equilibrium temperatures, and measured atmospheric metallicities to retrieve planetary bulk metallicities and core masses. Following results on the giant planets in the solar system and recent development in planet formation, we implement two interior structure models: one with a homogeneous envelope and one with an inhomogeneous one. Our method is first evaluated using a test planet and then applied to a sample of 37 giant exoplanets with observed atmospheric metallicities from the pre-JWST era. Although neither internal structure model is preferred with the current data, it is possible to obtain information on the interior properties of the planets, such as the core mass, through atmospheric measurements in both cases. We present updated metal mass fractions, in agreement with recent results on giant planets in the solar system.
\end{abstract}


\section{Introduction}
\label{sec:intro}

The amount of metals in an exoplanet has long been considered an indicator of the relative accretion of gas and solids during planet formation, as well as an important tracer of the location where the planets are born \citep{venturini2016}. With most of the planet's metal mass contained in its interior, finding links between metallicities observed in exoplanet atmospheres and interior properties is essential for the interpretation of new results coming from JWST.

Due to a lack of constraints, previous papers modelling the interior structure of exoplanets used simple models, in which the interior of the planets is made of a core of heavy elements, surrounded by a homogeneous, enriched envelope \citep{thorngren2016, muller}. The metallicity of this envelope is assumed to be the same metallicity as observed in their atmospheres. Nevertheless, recent results for Jupiter using data from the Juno mission \citep{wahl2017, netelmann2021, miguel2022} and on Saturn using ring seismology \citep{mankovich2021}, show that giant planet interiors are more complex than previously thought. All these models show inhomogeneous envelopes with a distribution of metals that gradually decreases from the core to the most external layer, a result that is also supported by the most recent formation models of giant planets \citep{lozovsky2018, valletta2022}.

Motivated by these results and by the extremely accurate atmospheric metallicities that we are expecting to find with measurements by JWST, in this paper, we explore the possibility of retrieving interior parameters of exoplanets using different models. For this, we adapt the static model developed for Jupiter by \citet{miguel2022} to exoplanets, considering masses, radii, atmospheric metallicities and helium fractions as prior data for a Bayesian fitting code. We retrieve bulk metallicities and core masses using both a homogeneous and an inhomogeneous distribution of heavy elements in the envelope. We combine evolution and static models to speed up the retrieval.

We describe the modelling of the interior and the fitting method used to estimate interior parameters in Section~\ref{sec:models}. In Section~\ref{sec:results_test}, we first apply this method to a test planet. We then apply it to a sample of real exoplanets in Section~\ref{sec:results_real}. In Section~\ref{sec:discuss}, we discuss our results, and Section~\ref{sec:concl} contains our conclusions.

\section{Methods}\label{sec:models}
\subsection{Planetary interior structure}
In this work, we use CEPAM to model the interiors of giant exoplanets. CEPAM \citep{guillot1995} was originally developed to model interiors of the giant planets in the solar system, but has also been used for exoplanets \citep[e.g][]{guillot2002}. The code has been developed further, and the last version includes a variety of atmospheric limits to be able to use it on inflated exoplanets \citep{parmentier2014, parmentier2016}, and includes different potential internal structures with homogeneous or inhomogeneous envelopes \citep{miguel2022}. This code uses the properties of the planet to calculate the observable parameters. For exoplanets, we use CEPAM to calculate the expected planetary radius. We use static modelling to model the planetary interiors throughout this work, but use evolution models to determine the range of potential luminosity values to use according to the age of each planet (see Section~\ref{sec:params}).
\begin{figure}
    \centering
    \includegraphics[width=0.5\columnwidth]{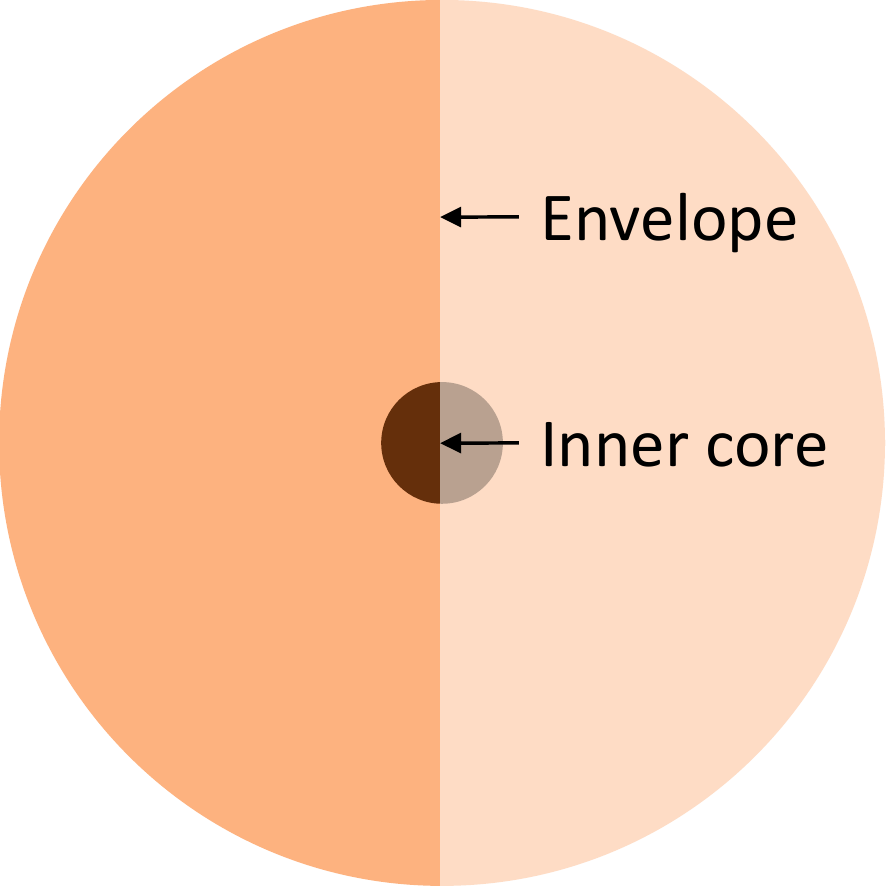}
    \caption{A schematic illustration of a model of the interior of a planet with a core of heavy elements and a homogeneous envelope.}
    \label{fig:layers_hom}
\end{figure}

\begin{figure}
    \centering
    \includegraphics[width=0.5\columnwidth]{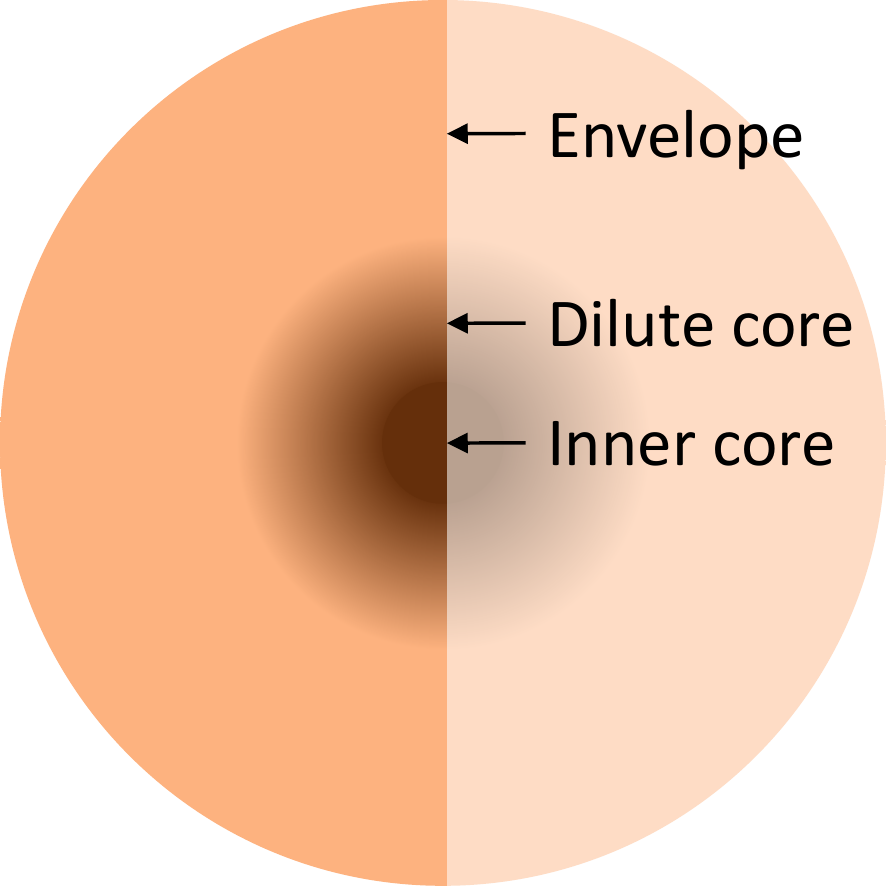}
    \caption{A schematic illustration of a model of the interior of a planet with an inhomogeneous envelope.}
    \label{fig:layers_dilute}
\end{figure}

The homogeneous model has a simple structure, and consists of a core made of heavy elements surrounded by an envelope with a homogeneous distribution of metals (see Figure \ref{fig:layers_hom}). In this model, the atmospheric metallicities observed are equal to the overall metal fraction in the giant planet's envelope.

On the other hand, models with inhomogeneous envelopes consist of a core and an envelope, where the heavy elements in the envelope are inhomogeneously distributed and gradually decrease from the core towards the atmosphere (see Figure \ref{fig:layers_dilute}). In this paper, we use the same formalism as in \cite{Guillot2018}, \citet{miguel2022} and \citet{Howard2023} to describe the increase of metals in the dilute core region based on the best fit to the observations. The metal fraction at each point in the envelope is described by,
\begin{equation}
\label{eq:metal_mass}
    Z(m)= Z_{\mathrm{atm}} + \frac{Z_{\mathrm{dilute}}-Z_{\mathrm{atm}}}{2} \left[1-\textrm{erf}\left(\frac{m - m_{\mathrm{dilute}}}{\delta m}\right)\right],
\end{equation}
where $Z(m)$ is the heavy element fraction at the mass coordinate $m$, $Z_{\mathrm{atm}}$ is the heavy element fraction observed in the atmosphere, which is effectively the outer boundary, $Z_{\mathrm{dilute}}$ is the maximal mass fraction of heavy elements in the envelope, $m_{\mathrm{dilute}}$ is the mass coordinate at which the metallicity gradient in the dilute core is steepest, and $\delta m$ is the slope of the gradual change in the heavy elements mass fraction. We set the parameter $\delta m$ equal to 0.075, which is the value used in \citet{miguel2022}. We find that varying $\delta m$ does not have a significant impact on the radius. Our tests show that the difference in radius between a run with $\delta m$ set to 0.075 and a run with $\delta m$ set to 0.25 is less than 0.3\%. 

Because giant exoplanets are made primarily of hydrogen and helium, the equations of state of these elements are very important to properly determine the interior structure \citep{Miguel2016,miguel2022}. In this work we use the MH13-H equation of state for hydrogen \citep{mh13, Miguel2016}, the SCVH95-He equation of state for helium \citep{scvh}, and the SESAME equation of state for water \citep{sesame} to describe the heavy elements, assumed to be ices, in the envelope. We assume that all the heavy elements in the envelope and atmosphere (including the dilute core region) are in ices, and that the small, compact core is made of pure rocks, with a structure based on \citet{Hubbard1989}.

\subsubsection{The atmospheric boundary}
In this work, we use a non-grey opacity model for a solar composition atmosphere, described in \citet{parmentier2014} and \citet{parmentier2016} and based on a classical model by \citet{Chandrasekhar1935}. The model relies on the equilibrium temperature (T$_{\mathrm{eq}}$) to include the effect of irradiation on the planetary structure. Following \citet{parmentier2014}, we fix the optical depth limit to $\tau_{\mathrm{lim}}=10^2$. We tested the impact of varying this parameter and found that it has a small impact on our calculations. The changes in radius caused by varying $\tau_{\mathrm{lim}}$ between $1$ and $10^{3}$ are $\lesssim$1\%. We set the other parameters to the default values recommended in \citet{parmentier2014} for giant planets. 

\subsection{Fitting routine}\label{sec:fitting}

To fit the interior structure model to the measurements, we use the algorithm \textsc{MultiNest} (v\,3.10) \citep{Feroz2013}, which implements nested sampling. We use the \texttt{Python} implementation described in \cite{2014A&A...564A.125B}. \textsc{MultiNest} uses Bayesian evidence to compare models.

Defining the data ${D}$ as the scalar of the measurement of the radius of the planet, the parameter vector $\boldsymbol{\theta}$ as the vector containing the values of the parameters of the model, and $M$ as the model itself, we can use Bayes' theorem
\begin{equation}
    \Pr(\boldsymbol{\theta}|{D}, M) = \frac{\Pr({D}|\boldsymbol{\theta}, M) \Pr(\boldsymbol{\theta}| M)}{\Pr({D}|M)} ,
\end{equation}
\noindent to determine $\Pr(\boldsymbol{\theta}|\boldsymbol{D}, M)$, the posterior distribution of the model parameters. The posterior distribution represents the updated belief of the model parameters given the data. $\Pr({D}|\boldsymbol{\theta}, M) \equiv \mathcal{L}(\boldsymbol{\theta})$ is the likelihood function. This is the likelihood of observing this set of data given the model and parameters. $\Pr(\boldsymbol{\theta}| M) \equiv \Pi(\boldsymbol{\theta})$ is the prior information for a model and $\Pr({D}|M) \equiv \mathrm{Z}$ is the Bayesian evidence, an indication of how well the model predicts the observed data. In this work, we use $\textrm{Z}_{\textrm{ev}}$ to indicate the evidence value, so as to not confuse it with the metal fractions.

CEPAM uses the properties of the planet to calculate the radius. For this reason, we use the radius in the likelihood function. We assume the uncertainty on the radius is normally distributed, with the mean being the measured value and the standard deviation equal to the uncertainty on the measurement. With this assumption, we can define the joint log-likelihood function as 

\begin{equation}
    \ln \mathcal{L}(\boldsymbol{\theta}) = - \frac{1}{2} \left[ \frac{(D - R_{\textrm{model}})^2}{\sigma^2} + \ln(2\pi \sigma^2) \right],
\end{equation}
\noindent where $D$ is the measured radius and $R_{\textrm{model}}$ is the predicted radius by the model $M$ with parameters $\boldsymbol{\theta}$. This log-likelihood function is then convolved with priors for each parameter, based on physical constraints and previous measurements. An example of a physical constraint is that the metal fraction of a planet cannot be less than zero.

In Bayes' theorem, Bayesian evidence is required to normalise the posterior over the volume of the prior. This is defined by
\begin{equation}
    \textrm{Z}_{\textrm{ev}} = \int \int ... \int \mathcal{L}(\boldsymbol{\theta}) \Pi(\boldsymbol{\theta}) d\boldsymbol{\theta},
\end{equation}
\noindent where the dimensionality of the integration is equal to the number of free parameters in the model.
The \textsc{MultiNest} algorithm initialises a number of live points sampling the prior space. It then contracts the distribution around points of high likelihood by discarding the points with the lowest likelihood and re-initialising them according to the prior distributions. This is repeated until the region of maximum likelihood is found.

The number of points the algorithm uses combined with the sampling efficiency determines how accurate the resulting parameters and evidence values are. \citet{Feroz2013} recommend using 400 points with a sampling efficiency of 0.3 for evidence calculation. We find that in this mode, the log evidence values are not always constant, but the value only varies by $\pm$2. Other modes perform equally well or worse. We therefore use the evidence estimation mode for all evidence calculations.

To assess which model best describes the data, we consider the difference in evidence values calculated by \textsc{MultiNest}. The best-fitting model between two competing models that are \textit{a priori} equally likely to describe the data is evaluated by considering the ratio of their evidence values. Expressed in log space, we write this as $\Delta \ln(\textrm{Z}_{\textrm{ev}}) = \ln(\textrm{Z}_2) - \ln(\textrm{Z}_1)$. Using an updated version of the Jefferys scale \citep[e.g.][]{Kass1995, Scaife2012, Callingham2015, Bloot2022}, $\Delta \ln(\textrm{Z}_{\textrm{ev}}) \ge 3$ is strong evidence that $M_2$ is significantly better at describing the data than $M_1$. $1<\Delta \ln(\textrm{Z}_{\textrm{ev}}) < 3$ is moderate evidence that $M_2$ describes the data better than $M_2$, and $0\le\Delta \ln(\textrm{Z}_{\textrm{ev}}) \le 1$ is inconclusive. However, in our case, we note that there is an extra uncertainty from the variation of the evidence value between runs. As mentioned before, the variation is at most $\pm$2, so we take $\Delta \ln(\textrm{Z}_{\textrm{ev}}) \ge 5$ as the condition for strong evidence that $M_2$ is significantly better at describing the data than $M_1$. We treat $\Delta \ln(\textrm{Z}_{\textrm{ev}}) \ge 3$ as a suggestion that $M_2$ is better at describing the data than $M_1$.

We note that this comparison requires that the two models are \textit{a priori} equally likely. If this is not the case, this comparison does not hold.

\subsubsection{Input and output parameters}\label{sec:params}
The input of the fitting method consists of the measured or inferred parameters, combined with any physical constraints. An example of such a constraint is that the core mass fraction cannot be less than 0.0, nor can it be more than 1.0.
The measured or inferred parameters used in this work are the mass, radius, metal fraction in the atmosphere, and equilibrium temperature. The radius is used to calculate the likelihood, whereas the mass, metal fraction in the atmosphere and equilibrium temperature are used as priors for the fitting method. For the mass and the equilibrium temperature, we use a Gaussian prior distribution centred on the measured or calculated value, with a width determined by their error bars. We use a logarithmic Gaussian prior distribution for the metal fraction in the atmosphere, as most metal fraction measurements are given in log space. The distribution is again centred on the measured value with a width determined by the error bars.

The physical constraints on the other parameters are determined in a few different ways. If the parameter is completely unconstrained, the input consists of the entire physically possible range. The unconstrained parameters in the default method are the core mass fraction and the parameters to determine the metal fraction in the envelope for the inhomogeneous case ($Z_{\mathrm{dilute}}$ and $m_{\mathrm{dilute}}$, see Equation~\ref{eq:metal_mass}). For these parameters, we use a prior uniformly distributed between 0.0 and 1.0. With these priors, it is technically possible to create a planet with a negative heavy element gradient. A negative metallicity gradient will be unstable, either succumbing to convection or a Rayleigh-Taylor instability, both of which will quickly restore a positive gradient. While these solutions are found to be very unlikely, they do occur occasionally. In those cases, we add an extra prior requiring that the gradient of heavy elements cannot be negative. For the helium fraction $Y$ and the heavy element fraction in the atmosphere $Z_{\mathrm{atm}}$, we define a constraint that $Y+Z_{\mathrm{atm}}\le 0.75$. This value was chosen to avoid a too large $Y+Z_{\mathrm{atm}}$ value in the envelope and to ensure convergence.
We use a Gaussian distribution for the helium fraction $Y$, centred on 0.277 with a standard deviation of 0.01.

The upper limit on the internal luminosity is set by running one homogeneous evolution model in CEPAM with all parameters set to the mean values of the measurements, with the helium fraction set to 0.3 and the core mass fraction set to 0.05. Out of all parameters, the one that has the largest impact on the internal luminosity in our tests is the metal fraction in the envelope. An increased atmospheric metallicity leads to an increased internal luminosity. Therefore, to calculate an upper limit on the luminosity, we set $Z_{\mathrm{atm}}$=0.5. We then select the highest value of the internal luminosity that agrees with the age of the star and use this as our upper limit.
The lower limit on the internal luminosity is obtained from \citet{guillot_gautier_2014} and \citet{sarkis2021}. We set it to $10^{24}\,\mathrm{erg\,s}^{-1}$, or 0.3\,$L_{\textrm{J}}$. 
If the upper limit on the internal luminosity is below or very close to this value, we set the lower limit to $10^{21}\, \mathrm{erg\,s}^{-1}$, $3\cdot 10^{-4}$\,$L_{\textrm{J}}$ instead. 
The range of the prior on the internal luminosity is fairly large. The goal of this is to not bias the fitting method to certain regions of the parameter space. In practice, this means that the calculated age of the best-fit set of parameters can be lower than the measured age, but still within the error bars. This is discussed in more detail in Section~\ref{sec:lum}.

The output of the model consists of the best-fit parameters and the log evidence value. Out of the best-fit parameters, the most relevant parameters to constrain the interior of a planet are the core mass fraction, the overall heavy element fraction of the planet, and the parameters to determine the metal fraction in the envelope for the inhomogeneous case.

\section{Results}
\subsection{Test planet}
\label{sec:results_test}
In this section, we test the robustness of the method by applying it to a simulated planet to analyse how well we can retrieve its interior structure. Our test planet is created using the properties in Table~\ref{tab:test_planet}.
\begin{table*}
    \centering
    \def\arraystretch{1.3}
    \begin{tabular}{|l|rrr|}
    \hline
    \hline
     & Test value & Run with only R, M & Run with R, M, T$_{\mathrm{eq}}$, $Z_{\mathrm{atm}}$, $Y$ \cr
     \hline
         Mass ($M_{\textrm{J}}$) & 0.478 & 0.478\,$\pm$0.02 & 0.478\,$\pm$0.02 \cr
         Core mass fraction & 0.0065 & [0.0, 1.0] & [0.0, 1.0] \cr
         $Y$ & 0.277& [0.0, 1.0] & 0.277\,$\pm$0.01 \cr
         $Z_{\textrm{atm}}$ & 0.11 & [0.0, 1.0] & -\cr
         log($Z_{\textrm{atm}})$ & -0.95 & - & -0.95\,$\pm$0.12\cr
         T$_{\mathrm{eq}}$ (K) & 1580 & [0.0, 6000] & 1585\,$\pm$24 \cr
         Internal luminosity ($L_{\textrm{J}}$) & 4.5 & [0.3, 5.8]& [0.3, 5.8]\cr
         m$_{\mathrm{dilute}}$ & 0.27 & [0.0, 1.0] & [0.0, 1.0] \cr
         $Z_{\mathrm{dilute}}$ & 0.2& [0.0, 1.0] & [0.0, 1.0]\cr
         Radius ($R_{\textrm{J}}$) & 1.5 & 1.5\,$\pm$0.06 & 1.5\,$\pm$0.06 \cr
    \hline
    \hline
    \end{tabular}
    \caption{Parameters used to create the test planet and the prior ranges used for the two runs. The test value is the value for that parameter used to create the planet. The first of the two runs only includes measurements of the radius R and the mass M, whereas the second run uses the radius R, mass M, equilibrium temperature T$_{\mathrm{eq}}$, metallicity of the atmosphere $Z_{\mathrm{atm}}$ and the helium fraction $Y$.
    The value of the internal luminosity used to create the test planet was chosen based on an evolution run, using an age of 1~Gyr. The prior ranges are written as $\mu \pm \sigma$ for the Gaussian distributions, where $\mu$ is the mean of the distribution and $\sigma$ is the standard deviation. For the uniform priors, the lower and upper limits of the uniform distribution are given. m$_{\mathrm{dilute}}$ is the mass coordinate where the dilute core region is located and $Z_{\mathrm{dilute}}$ is the dilute core central metal fraction.}
    \label{tab:test_planet}
\end{table*}

\subsubsection{Required measurements to constrain the interior structure}
\begin{figure*}
     \centering
     \begin{subfigure}
         \centering
         \includegraphics[width=0.45\textwidth]{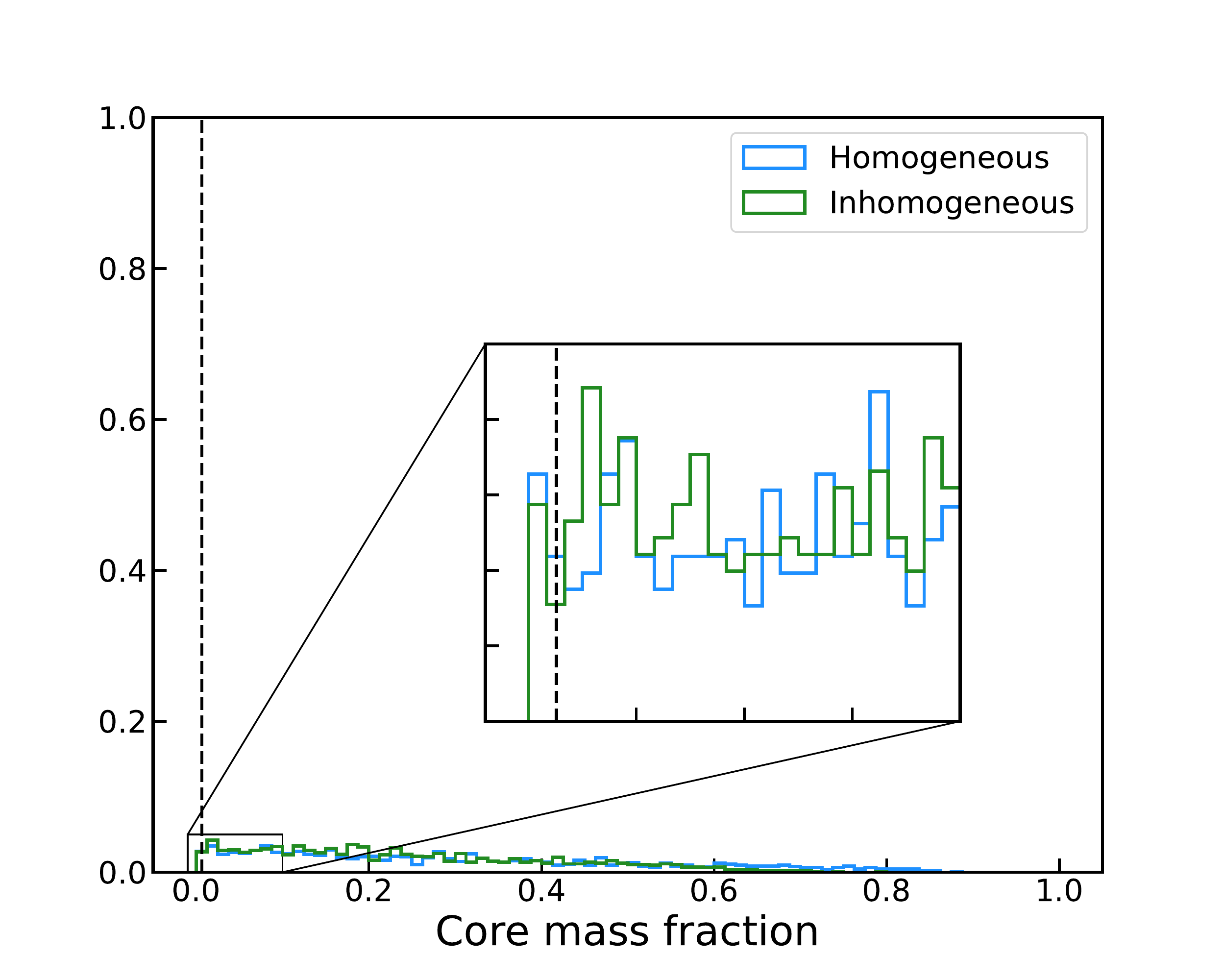}
         \label{fig:core_mr}
     \end{subfigure}\begin{subfigure}
         \centering
         \includegraphics[width=0.45\textwidth]{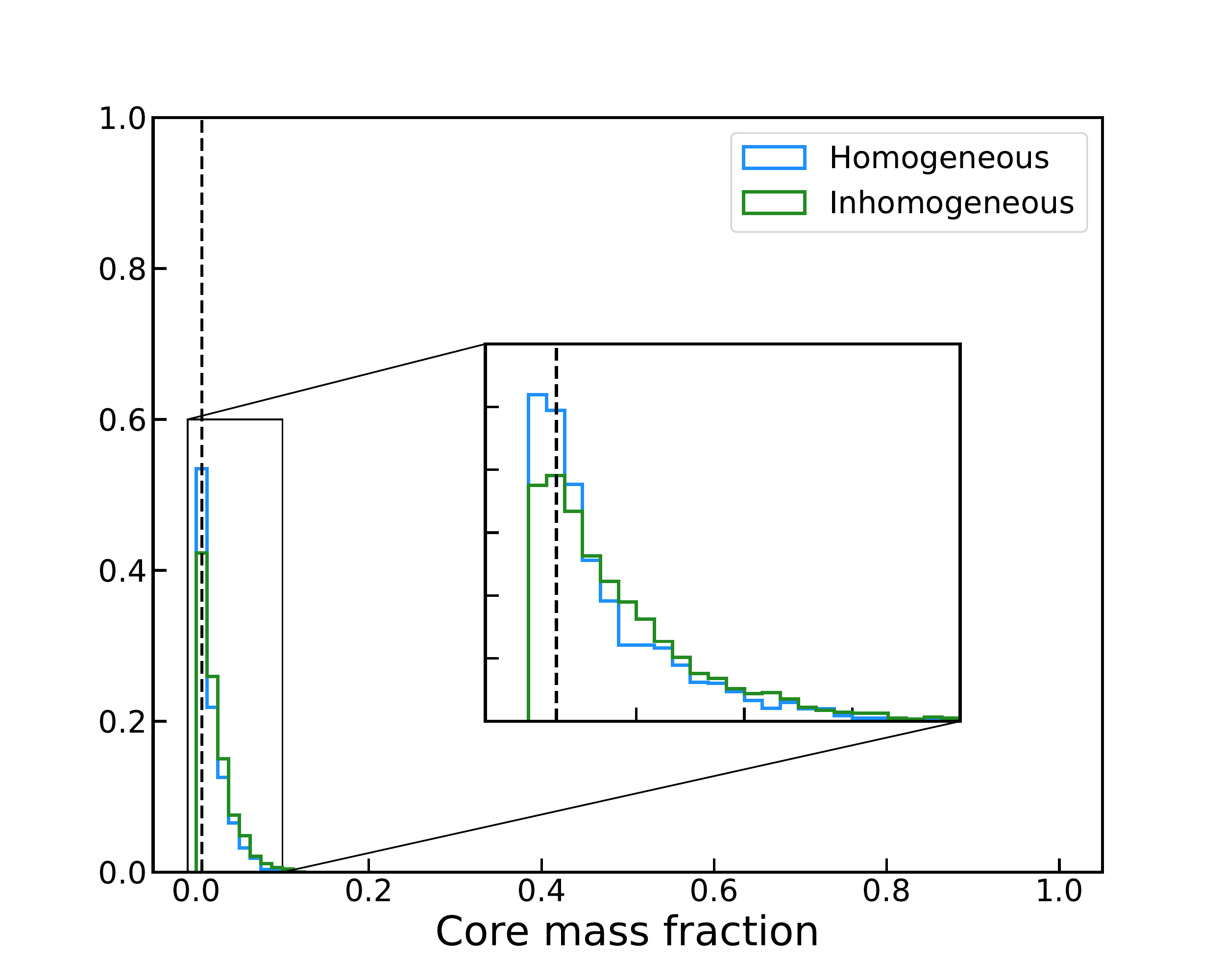}
         \label{fig:core_y}
     \end{subfigure}
        \caption{Histogram showing the distribution of the core mass fraction produced by fitting the two different interior structures to measurements of the radius and the mass (left panel) and the radius, mass, equilibrium temperature, helium fraction and atmospheric metal fraction (right panel) of the test planet. The light blue line shows the distribution of the core mass fraction for the homogeneous model fit, and the dark green line shows the distribution for the inhomogeneous model. The insets show the same distribution between 0.0 and 0.15 in more detail, using more bins. The dashed black line shows the core mass fraction used to create the test planet.}
        \label{fig:core_all}
\end{figure*}

In order to find the best-fitting model that reproduces the properties of the test planet, we check what parameters are necessary to break the degeneracies. The first test is to use only the mass and the radius of the planet as input for our method, and test if that information alone is enough to retrieve the interior structure of the test planet. For the radius, we use a value of 1.5\,$\pm$0.06\,$R_{\textrm{J}}$, and we set the mass value to 0.478\,$\pm$0.02\,$M_{\textrm{J}}$, where the error bars are chosen in agreement to typical exoplanet measurement estimations, examples of which are shown in Table~\ref{tab:planets}. With just these constraints, we find that the method struggles to constrain any of the parameters. This is shown in Figure~\ref{fig:core_all} (left panel), where we plot the histogram of the core mass fraction distribution that we find for both models.
We see that the core mass fraction is completely unconstrained in these runs. The core mass fraction in the homogeneous model is $0.262^{+0.296}_{-0.190}$. For the inhomogeneous model, it is $0.211^{+0.230}_{-0.146}$. The same is found for the other unconstrained parameters. 

We observe that in this case, where we use only the mass and radius of the planet as prior data, we cannot retrieve the true core mass fraction of the planet. This can be seen in Figure \,\ref{fig:core_all} (left panel), where some of the extreme values retrieved for the core mass fraction are close to 0.8, while the true value of this parameter in the test planet is 0.0065. This is because, in a system with few constraints and many unknowns, the problem is degenerate and the code can obtain a solution that is far from the true value. For example, when the core mass fraction is high, other parameters, such as the metallicity and helium fraction of the atmosphere, as well as the internal luminosity, can compensate for the effect of the large core to still create a planet with the desired radius. This is therefore a valid solution for the system, even if it is not close to the right one.

It is clear that the method requires more information to find a good fit. This is expected, as a fitting method generally requires more constraints than free parameters to converge to a solution. When only using the mass and the radius as input, we have 4 free parameters in the homogeneous model and 6 in the inhomogeneous model. In this case, the 2 constraints are not enough to allow the method to converge.

Other parameters that can be determined observationally are the atmospheric metal fraction and the equilibrium temperature (via the determination of the stellar parameters and planetary albedo). Moreover, we can also add the helium fraction as a constraint in the calculations. While helium is a challenging element to be spectroscopically detected in stars or exoplanet atmospheres, one of the approaches typically followed by stellar modellers is to determine the initial helium abundance by setting free values of $Y_{ini,star}$ in a range that is in reasonably good agreement with solar values \citep{Nsamba2021}. We can then assume that the exoplanets have the same helium fraction as their primitive host star, an assumption also made for the overall helium abundance of the giant planets in the solar system.

To test the effect of adding these measurements, we run the method again on the same test planet, with the atmospheric metal fraction, the equilibrium temperature and the helium fraction added as input priors. The prior parameters are listed in Table~\ref{tab:test_planet}.
The uncertainties on these values are based on uncertainties of measurements of real exoplanets, examples of which can be found in Table~\ref{tab:planets}.
When we apply the method using these constraints, we find a distribution in core mass fractions as shown in Figure~\ref{fig:core_all} (right panel). The core mass is much more constrained in these runs, as we can now see that the core mass is less than 0.1, independent of the interior structure. For the homogeneous model, the core mass is now $0.012^{+0.021}_{-0.008}$. The inhomogeneous model is again very similar, with the core mass being $0.015^{+0.023}_{-0.011}$. This last case using the equilibrium temperature, atmospheric metal fraction and helium fraction as input parameters in addition to mass and radius is the best case to retrieve the interior properties of the planets and the one that will be used during the rest of this manuscript.

\subsubsection{Internal luminosity}
\label{sec:lum}

\begin{figure}
    \centering
    \includegraphics[width=0.95\columnwidth]{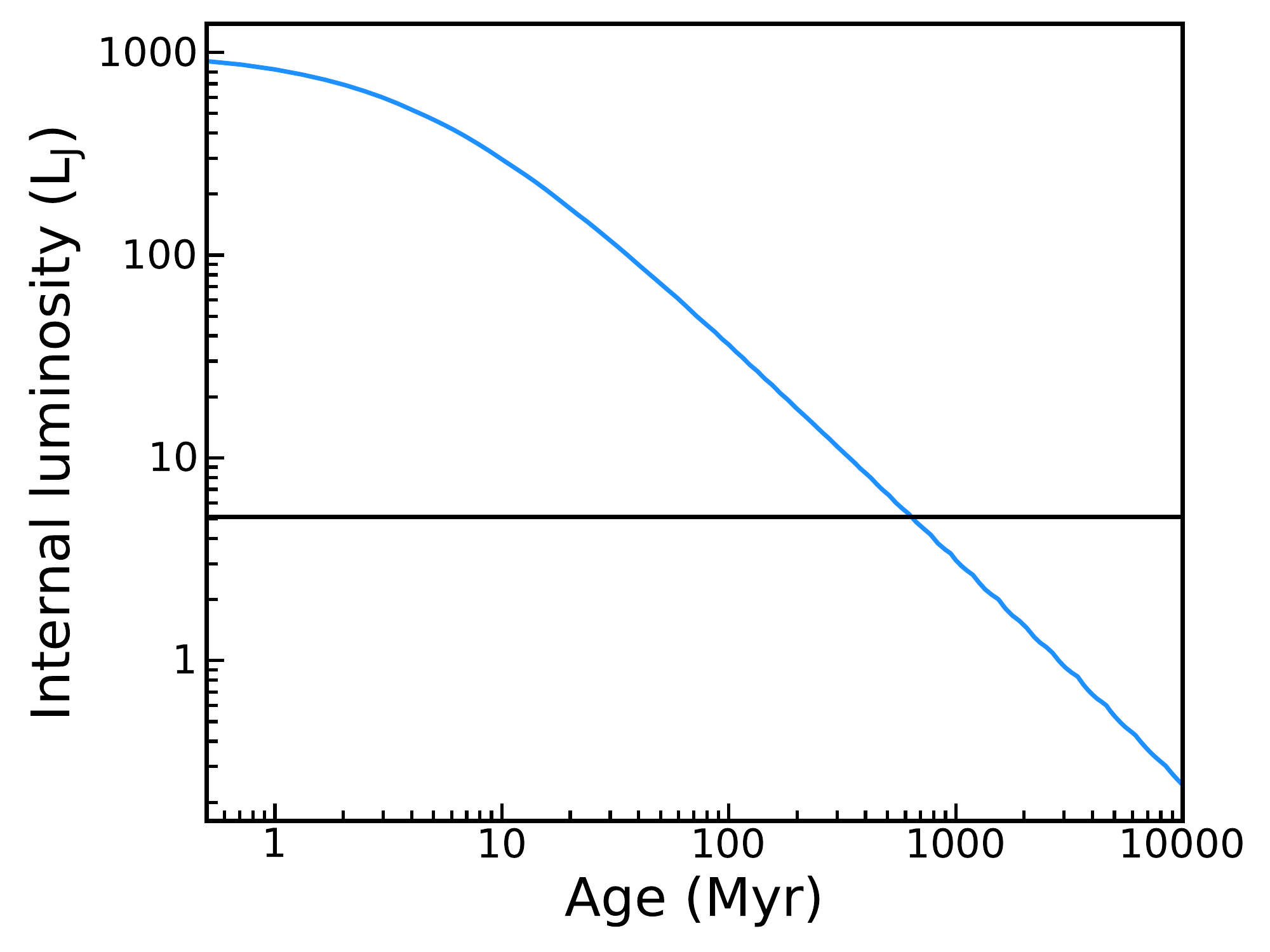}
    \caption{Evolution of the internal luminosity as a function of time for the test planet. The blue line shows the results from an evolution run using the best-fit parameters for the test planet, using the homogeneous model with a helium fraction constraint. The black line indicates the best-fit internal luminosity calculated with the fitting method using static modelling. The point where the two lines cross is at 622\,Myr.}
    \label{fig:ev_line}
\end{figure}

In Section~\ref{sec:params}, we describe the upper limit on the internal luminosity, based on the age of the star used in an evolution model of the planet. To test the validity of this assumption for the upper limit, we run the evolution model for the best-fit parameters found with the fitting method and static model for the test planet and compare this with the internal luminosity found with this fitting routine. The results of this run are plotted in Figure~\ref{fig:ev_line}. We compare the results of this run, particularly the radius and the internal luminosity, to the radius and luminosity found by the static runs using the upper limit on the internal luminosity. We find that the internal luminosity found by the static runs is realistic when looking at the full evolution run. The age of the planet that agrees with both the calculated radius and internal luminosity is 0.6\,Gyr. The age chosen to create the planet is a little higher at 1\,Gyr, but well within typical observational error bars when compared to the values in Table~\ref{tab:planets}.

\subsubsection{Homogeneous versus inhomogeneous structures}
A consequence of using this fitting method is that it produces an evidence value for each run. Comparing these evidence values could give us information about which model describes the data better.
We list the found log evidence values for our simulated planet in Table~\ref{tab:evs_test_planet}. 
We see that there is no significant difference between the evidence values of the two models. This indicates that the parameter space is still too large for the fitting routine to find a unique solution. There are simply too many degeneracies to find one single solution. More constraints would be needed to better determine which one of these two models fits exoplanet data better and would be the study of future research.

\begin{table}
    \centering
    \def\arraystretch{1.3}
    \begin{tabular}{|l|r|}
         \hline
         \hline
         Run & ln(Z$_{\textrm{ev}}$) \cr
         \hline
         Homogeneous & -33.1 $\pm$ 0.05 \cr
         Inhomogeneous & -32.3 $\pm$ 0.2 \cr
         \hline
         \hline
    \end{tabular}
    \caption{The log evidence values found for the test planet for both models, with constraints on R, M, $Z_{\mathrm{atm}}$, T$_{\mathrm{eq}}$ and $Y$.}
    \label{tab:evs_test_planet}
\end{table}

\begin{figure*}
    \centering
    \begin{subfigure}
         \centering
         \includegraphics[width=0.45\textwidth]{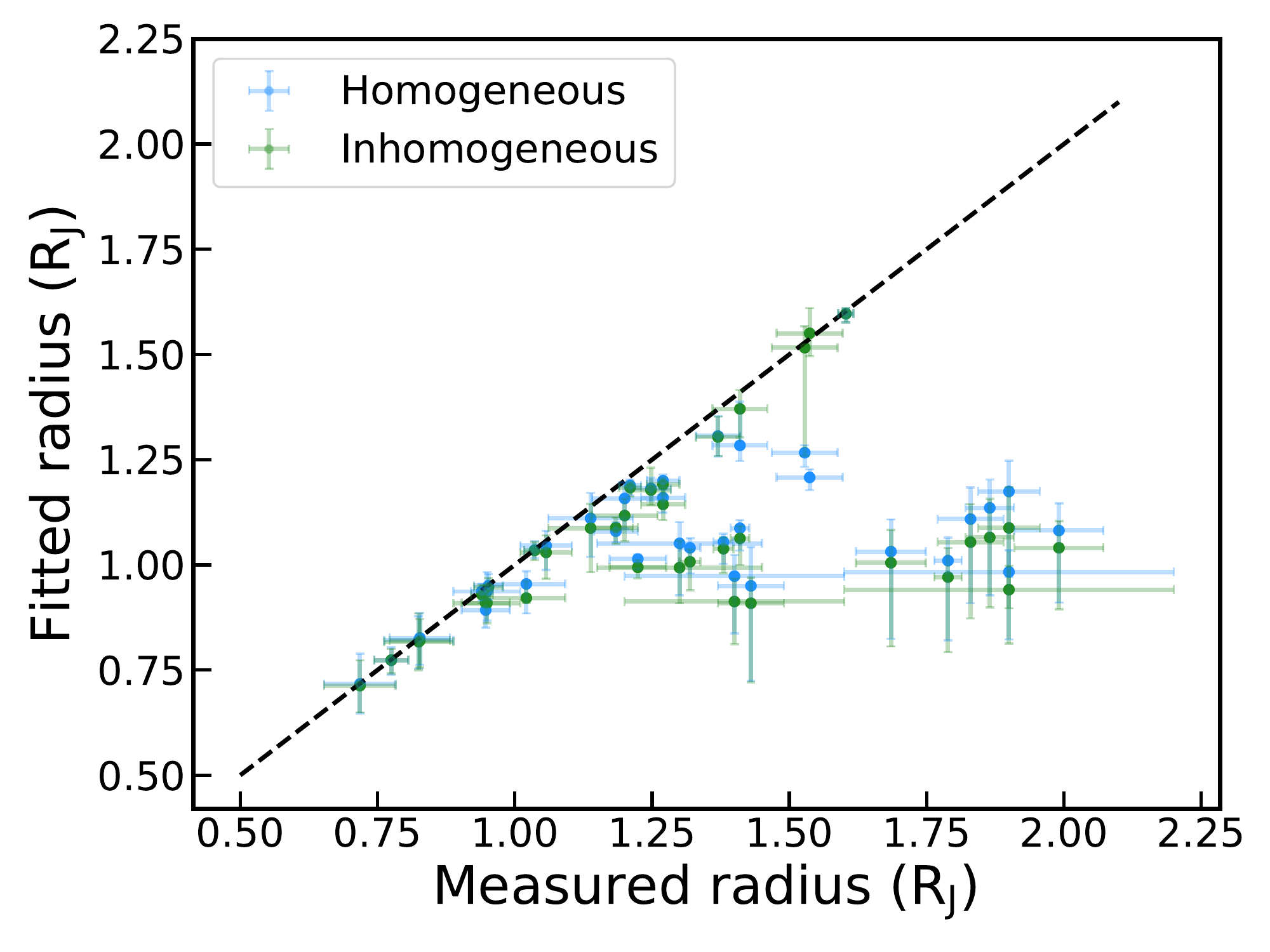}
    \end{subfigure}\begin{subfigure}
         \centering
         \includegraphics[width=0.45\textwidth]{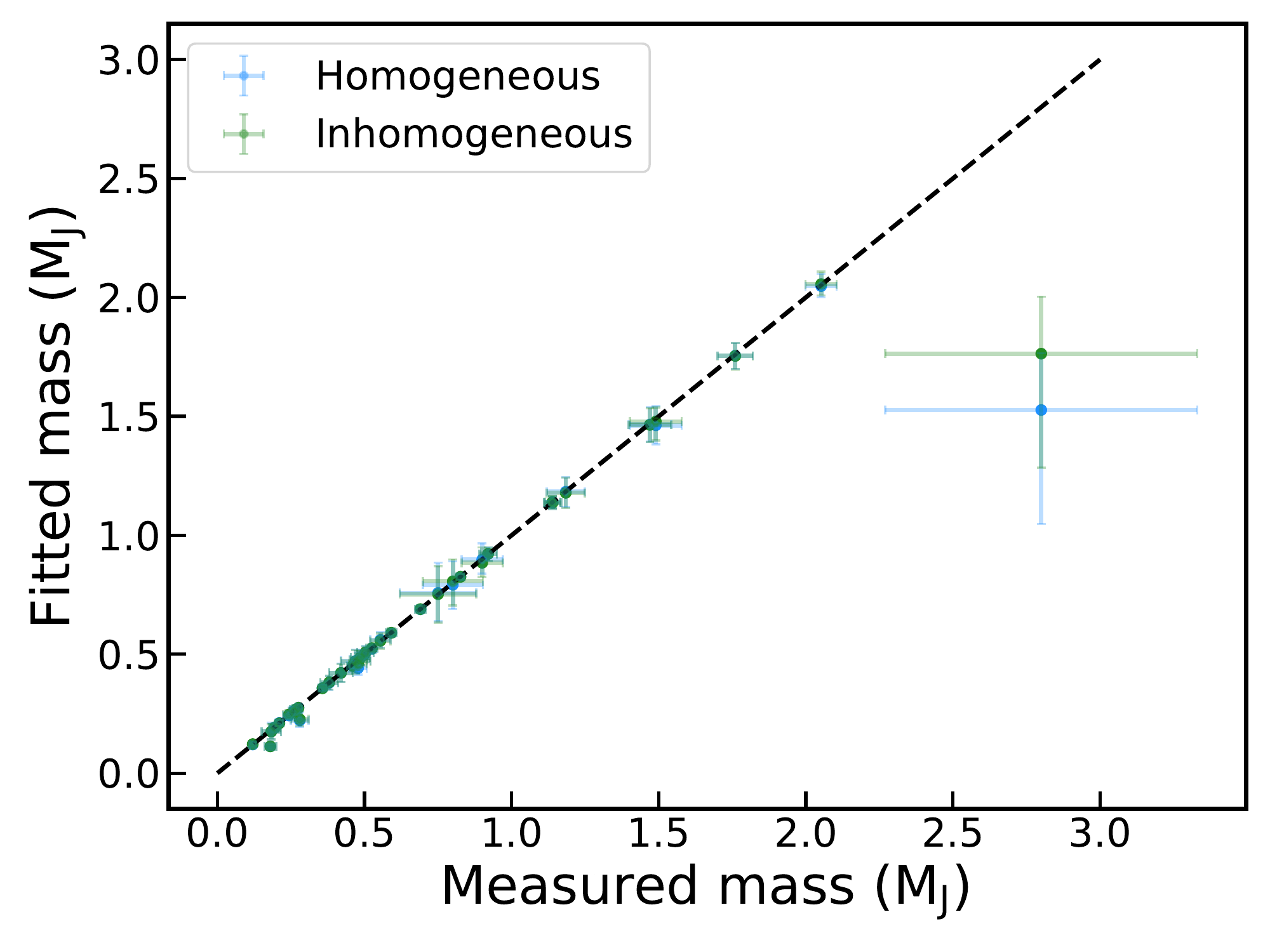}  
    \end{subfigure}     
    \caption{Comparison between measured and fitted parameters. Left panel: Measured radius --used as input for the fitting method-- on the horizontal axis, and the radius found by the fitting method on the vertical axis. Right panel: measured mass --used as input for the fitting method-- on the horizontal axis, and the mass found by the fitting method on the vertical axis. In both figures, the light blue and dark green points respectively correspond to the homogeneous and inhomogeneous models. A line showing where they are equal is drawn. The error bars represent the 1$\sigma$ uncertainty.}
    \label{fig:rad}
\end{figure*}
\subsection{Sample of exoplanets}
\label{sec:results_real}
After analysing the robustness of our method with a test planet, we apply it to real exoplanets. In this section, we present the results of retrieving interior parameters for a sample of 37 exoplanets. These planets were selected based on several criteria. First, they need to have available measurements of the radius and the mass, as well as an inferred equilibrium temperature. Because we use atmospheric water abundances as an estimate for the metal fraction in the atmosphere, we select planets with measured water abundances. Finally, our planets are in a mass range of $0.1 M_{\textrm{J}}$ to $3 M_{\textrm{J}}$. The entire selection of exoplanets with the corresponding measured values used to constrain the calculations is listed in Table~\ref{tab:planets}, where we also include stellar ages when available. We note that when stellar age is not available, we use 1\,Gyr as the age of the planet for the determination of the upper limit on its internal luminosity.

The results of all runs, including corner plots of the distributions, are available at \url{https://github.com/AstroYamila-Team/exoplanet-interior-retrievals}.

\begin{table*}
\def\arraystretch{1.4}
    \centering
    \begin{tabular}{|l|rrrrr|l|}

    \hline
    \hline
Name & Mass ($M_{\textrm{J}}$) & Radius ($R_{\textrm{J}}$) & T$_{\mathrm{eq}}$ (K) & log(X$_{H_2O}$) & Stellar age (Gyr) & References \cr
\hline

51 Peg b $\star$& $0.47^{+0.03} _{-0.07}$ & $1.9^{+0.3} _{-0.3}$ & $1226^{+72} _{-69}$ & $-3.5\pm 1^*$ & $4^{+2.5} _{-2.5}$ & 1, 2, 3 \cr
HAT-P-1 b $\star$ & $0.525^{+0.019} _{-0.019}$ & $1.319^{+0.019} _{-0.019}$ & $1322^{+14} _{-15}$ & $-2.72^{+0.42} _{-0.52}$ & $3.6$ & 4, 5, 6 \cr
HAT-P-12 b & $0.2105^{+0.013} _{-0.013}$ & $0.949^{+0.061} _{-0.061}$ & $957\pm 20$ & $-3.91^{+1.01} _{-1.89}$ & $2.5^{+2} _{-2}$ & 4, 7, 8, 9 \cr
HAT-P-18 b & $0.183^{+0.034} _{-0.032}$ & $0.947^{+0.044} _{-0.044}$ & $841\pm 15$ & $-2.63\pm 1.18$ & $12.4^{+6.4} _{-6.4}$ & 10, 11, 12, 13 \cr
HAT-P-3 b & $0.591^{+0.018} _{-0.018}$ & $0.827^{+0.055} _{-0.055}$ & $1127^{+49} _{-39}$ & $-6.93\pm 2.73$ & $1.6^{+1.3} _{-1.3}$ & 11, 14, 15 \cr
HAT-P-32 b $\star$& $0.75^{+0.13} _{-0.13}$ & $1.789^{+0.025} _{-0.025}$ & $1786\pm 26$ & $-2.84\pm 0.92$ & $3.8^{+0.5} _{-0.5}$ & 11, 16, 17 \cr
HAT-P-38 b & $0.267^{+0.02} _{-0.02}$ & $0.825^{+0.063} _{-0.063}$ & $1082\pm 55$ & $-4.29\pm 2.16$ & $10.1^{+4.8} _{-4.8}$ & 11, 18\cr
HAT-P-41 b $\star$ & $0.8^{+0.102} _{-0.102}$ & $1.685^{+0.076} _{-0.051}$ & $1941\pm 38$ & $-2.77\pm 1.09$ & $2.2^{+0.4} _{-0.4}$ & 11, 19 \cr
HD 149026 b & $0.357^{+0.014} _{-0.011}$ & $0.718^{+0.065} _{-0.065}$ & $1634^{+90} _{-23}$ & $-5.75\pm 2.91$ & $2^{+0.8} _{-0.8}$ & 11, 14, 20, 21, 22\cr
HD 189733 b & $1.138^{+0.025} _{-0.025}$ & $1.138^{+0.077} _{-0.077}$ & $1201^{+13} _{-12}$ & $-5.04^{+0.46} _{-0.30}$ & $0.6$ & 4, 14, 23, 24 \cr
HD 209458 b $\star$ & $0.69^{+0.017} _{-0.017}$ & $1.38^{+0.018} _{-0.018}$ & $1449^{+12} _{-12}$ & $-4.66^{+0.39} _{-0.30}$ & $4^{+2} _{-2}$ & 4, 14, 24, 25 \cr
KELT-11 b & $0.195^{+0.019} _{-0.019}$ & $1.3^{+0.18} _{-0.12}$ & $1712^{+51} _{-46}$ & $-4.73^{+1.13} _{-1.51}$ & & 26, 27 \cr
WASP-101 b & $0.5^{+0.004} _{-0.004}$ & $1.41^{+0.05} _{-0.05}$ & $1559\pm 38$ & $-6.95\pm 2.61$ & $0.9^{+0.4} _{-1.3}$ & 11, 28, 29 \cr
WASP-103 b & $1.49^{+0.088} _{-0.088}$ & $1.528^{+0.073} _{-0.047}$ & $2513\pm 49$ & $-1.73^{+0.38} _{-0.55}$ & $4^{+1} _{-1}$ & 28, 30, 31 \cr
WASP-107 b & $0.12^{+0.01} _{-0.01}$ & $0.94^{+0.02} _{-0.02}$ & $770\pm 60$ & $-1.7^{+0.30} _{-0.60}$ & & 32, 33 \cr
WASP-117 b & $0.2755^{+0.009} _{-0.009}$ & $1.021^{+0.065} _{-0.076}$ & $1225^{+36} _{-39}$ & $-3.82^{+1.37} _{-1.55}$ & $4.6^{+2} _{-2}$ & 34,35 \cr
WASP-12 b $\star$& $1.47^{+0.076} _{-0.069}$ & $1.9^{+0.057} _{-0.055}$ & $2546\pm 82$ & $-3.16^{+0.66} _{-0.69}$ & $1.7^{+0.8} _{-0.8}$ & 4, 28, 36 \cr
WASP-121 b $\star$& $1.184^{+0.065} _{-0.064}$ & $1.865^{+0.044} _{-0.044}$ & $2358\pm 52$ & $-3.05\pm 0.87$ & & 11, 37 \cr
WASP-127 b & $0.18^{+0.02} _{-0.02}$ & $1.37^{+0.04} _{-0.04}$ & $1404\pm 29$ & $-2.56^{+0.92} _{-4.65}$ & $11.41^{+1.8} _{-1.8}$ & 38, 39, 40 \cr
WASP-17 b $\star$ & $0.486^{+0.032} _{-0.032}$ & $1.991^{+0.081} _{-0.081}$ & $1771\pm 35$ & $-4.04^{+0.91} _{-0.42}$ & $3^{+2.6} _{-2.6}$ & 4, 41, 42 \cr
WASP-19 b $\star$ & $1.139^{+0.03} _{-0.03}$ & $1.41^{+0.017} _{-0.017}$ & $2099\pm 39$ & $-3.9^{+0.95} _{-1.16}$ & $11.5^{+2.7} _{-2.7}$ & 4, 28, 43, 44 \cr
WASP-29 b & $0.243^{+0.02} _{-0.02}$ & $0.775^{+0.031} _{-0.031}$ & $970^{+32} _{-31}$ & $-7.93\pm 2.38$ & $15^{+8} _{-8}$ & 11, 45, 46, 47\cr
WASP-31 b & $0.478^{+0.03} _{-0.03}$ & $1.537^{+0.06} _{-0.06}$ & $1575\pm 32$ & $-3.97^{+1.01} _{-2.27}$ & & 4, 48\cr
WASP-33 b & $2.8^{+0.53} _{-0.53}$ & $1.603^{+0.014} _{-0.014}$ & $2781\pm 41$ & $-6.8^{+3.00} _{-4.00}$ & & 49, 50, 51 \cr
WASP-39 b & $0.28^{+0.03} _{-0.03}$ & $1.27^{+0.04} _{-0.04}$ & $1030^{+30} _{-20}$ & $-1.37^{+0.05} _{-0.13}$ & & 4, 52, 53 \cr
WASP-43 b & $2.052^{+0.053} _{-0.053}$ & $1.036^{+0.019} _{-0.019}$ & $1444\pm 40$ & $-4.36\pm 2.1$ & $0.4$ & 11, 28, 54 \cr
WASP-52 b & $0.46^{+0.02} _{-0.02}$ & $1.27^{+0.03} _{-0.03}$ & $1315\pm 35$ & $-4.09\pm 0.87$ & $0.4^{+0.3} _{-0.3}$ & 15, 55\cr
WASP-6 b & $0.503^{+0.019} _{-0.038}$ & $1.224^{+0.051} _{-0.052}$ & $1194^{+58} _{-57}$ & $-6.91^{+1.83} _{-2.07}$ & $11^{+7} _{-7}$ & 4, 56 \cr
WASP-63 b $\star$& $0.38^{+0.03} _{-0.03}$ & $1.43^{+0.06} _{-0.06}$ & $1536\pm 37$ & $-5.81\pm 2.81$ & & 11, 28, 57 \cr
WASP-67 b & $0.42^{+0.04} _{-0.04}$ & $1.4^{+0.2} _{-0.2}$ & $1040\pm 30$ & $-5.61^{+1.98} _{-0.95}$ & & 11, 57, 58 \cr
WASP-69 b & $0.26^{+0.017} _{-0.017}$ & $1.057^{+0.047} _{-0.047}$ & $963\pm 18$ & $-3.94\pm 1.25$ & $2$ & 11, 59 \cr
WASP-74 b & $0.826^{+0.014} _{-0.014}$ & $1.248^{+0.036} _{-0.036}$ & $1922\pm 46$ & $-5.91\pm 2.81$ & $2^{+1} _{-1.6}$ & 11, 28, 59, 60, 61 \cr
WASP-76 b $\star$& $0.92^{+0.03} _{-0.03}$ & $1.83^{+0.06} _{-0.06}$ & $2190\pm 43$ & $-2.7\pm 1.07$ & & 11, 28, 62 \cr
WASP-77 A b & $1.76^{+0.06} _{-0.06}$ & $1.21^{+0.02} _{-0.02}$ & $1715^{+26} _{-25}$ & $-3.93^{+0.10} _{-0.09}$ & $1^{+0.3} _{-0.5}$ & 63, 64, 65 \cr
WASP-80 b & $0.554^{+0.03} _{-0.039}$ & $0.952^{+0.026} _{-0.027}$ & $825\pm 20$ & $-5.34\pm 2.65$ & & 11, 66 \cr
WASP-96 b & $0.48^{+0.03} _{-0.03}$ & $1.2^{+0.06} _{-0.06}$ & $1285\pm 40$ & $-3.65^{+0.90} _{-0.94}$ & $8^{+8} _{-26}$ & 29, 67 \cr
XO-1 b & $0.9^{+0.07} _{-0.07}$ & $1.184^{+0.04} _{-0.04}$ & $1204\pm 11$ & $-2.75\pm 1.64$ & $4.5^{+2} _{-2}$ & 68, 69, 70 \cr
\hline
\hline
\end{tabular}
\caption{Exoplanets used in this work. The planets marked with $\star$ are planets that are not well reproduced by our method and that are removed from consideration for the analysis of the results. $*$: There is no given uncertainty on the water abundance of 51~Peg~b, therefore we use a typical value based on the other exoplanets in the sample. References: 1: \citet{2017AJ....153..138B}, 2: \citet{2013ApJ...767...27B}, 3: \citet{2015A&A...576A.134M}, 
    4: \citet{pinhas}, 5: \citet{2014MNRAS.437...46N}, 6: \citet{2007ApJ...656..552B}, 
    7: \citet{Line_2013}, 8: \citet{Todorov_2013}, 9: \citet{2009ApJ...706..785H}, 
    10: \citet{refId0}, 11: \citet{Tsiaras_2018}, 12: \citet{2014ApJ...785..126K}, 13: \citet{2011ApJ...726...52H},
    14: \citet{2008ApJ...677.1324T}, 15: \citet{2011AJ....141..179C}, 
    16: \citet{2011ApJ...742...59H}, 17: \citet{2017A&A...602A.107B}, 
    18: \citet{2012PASJ...64...97S}, 
    19: \citet{2012AJ....144..139H}, 
    20: \citet{Sato_2005}, 21: \citet{2010A&A...524A..25T}, 22: \citet{2010MNRAS.408.1689S}, 
    23: \citet{2021AJ....162..176P}, 24: \citet{2006A&A...460..251M}, 
    25: \citet{2011MNRAS.418.1822W}, 
    26: \citet{Col_n_2020}, 27: \citet{Pepper_2017}, 
    28: \citet{2021ApJ...923..242G}, 29: \citet{2014MNRAS.440.1982H}, 
    30: \citet{2014A&A...562L...3G}, 31: \citet{2020MNRAS.497.5155W}, 
    32: \citet{2017A&A...604A.110A}, 33: \citet{2018Natur.557...68S}, 
    34: \citet{2014A&A...568A..81L}, 35: \citet{Anisman_2020}, 
    36: \citet{2017AJ....153...78C}, 
    37: \citet{2016MNRAS.458.4025D}, 
    38: \citet{2018A&A...616A.145C}, 39: \citet{2021AJ....162...36W}, 40: \citet{2017A&A...599A...3L}, 
    41: \citet{2011MNRAS.416.2108A}, 42: \citet{2010ApJ...709..159A}, 
    43: \citet{2013MNRAS.436....2M}, 44: \citet{2010ApJ...708..224H} 
    45: \citet{2013MNRAS.428.3680G}, 46: \citet{2021AJ....162..221S}, 47: \citet{2010ApJ...723L..60H}, 
    48: \citet{2011A&A...531A..60A}, 
    49: \citet{2015A&A...578L...4L}, 50: \citet{Chakrabarty_2019}, 51: \citet{2019A&A...622A..71V}, 
    52: \citet{2011A&A...531A..40F}, 53: \citet{Wakeford_2017}, 
    54: \citet{2012A&A...542A...4G}, 
    55: \citet{2013A&A...549A.134H}, 
    56: \citet{2009A&A...501..785G}, 
    57: \citet{2012MNRAS.426..739H}, 
    58: \citet{Bruno_2018}, 
    59: \citet{2014MNRAS.445.1114A}, 
    60: \citet{2020arXiv200711851L}, 61: \citet{2019MNRAS.485.5168M}, 
    62: \citet{2016A&A...585A.126W}, 
    63: \citet{2013PASP..125...48M}, 64: \citet{2021Natur.598..580L}, 65: \citet{2020A&A...636A..98C}, 
    66: \citet{2013A&A...551A..80T}, 
    67: \citet{Yip_2020}, 
    68: \citet{2006ApJ...648.1228M}, 69: \citet{2006ApJ...652.1715H}, 70: \citet{2018MNRAS.481.4261S}. 
    }
    \label{tab:planets}
    
    \end{table*}

\subsubsection{Reproducing the measurements}
To test if our method works well for the entire sample of exoplanets, a first check is to compare the radii and masses measured against the best-fit radii and masses found with the fitting routine. 
This comparison is shown in Figure~\ref{fig:rad}. 
The plot shows that most of the planets in the sample are well reproduced. For the masses, (Figure~\ref{fig:rad}, right panel) all the planets have masses that are well-fitted, with the exception of the most massive planet in the sample. We note, however, that the difference between the best-fit mass and the measured mass for this massive planet is less than 2$\sigma$, showing that the method can reproduce the masses very well.
However, when looking at the radius (Figure~\ref{fig:rad}, left panel) there is also a number of planets that are not properly fitted. These planets have a best-fit radius that is smaller than the measured radius. The difference between the best-fit radius and the measured radius for most inflated planets can be between 10 to 20$\sigma$.
The large difference between the best-fit radius and the measured radius is caused by the planet being very inflated. Due to the effect of stellar irradiation on exoplanets, there is extra energy deposited in the interior of the planet \citep{MolLous2020}. This might be another uncertainty and extra parameter to consider in our calculations for some extreme cases. We come back to these extreme planets in Section~\ref{sec:inflated}. In the rest of this work, we remove these planets from the sample, leading to a final sample of 25 exoplanets.

\subsubsection{Core mass distribution}
\begin{figure}
    \centering
    \includegraphics[width=0.95\columnwidth]{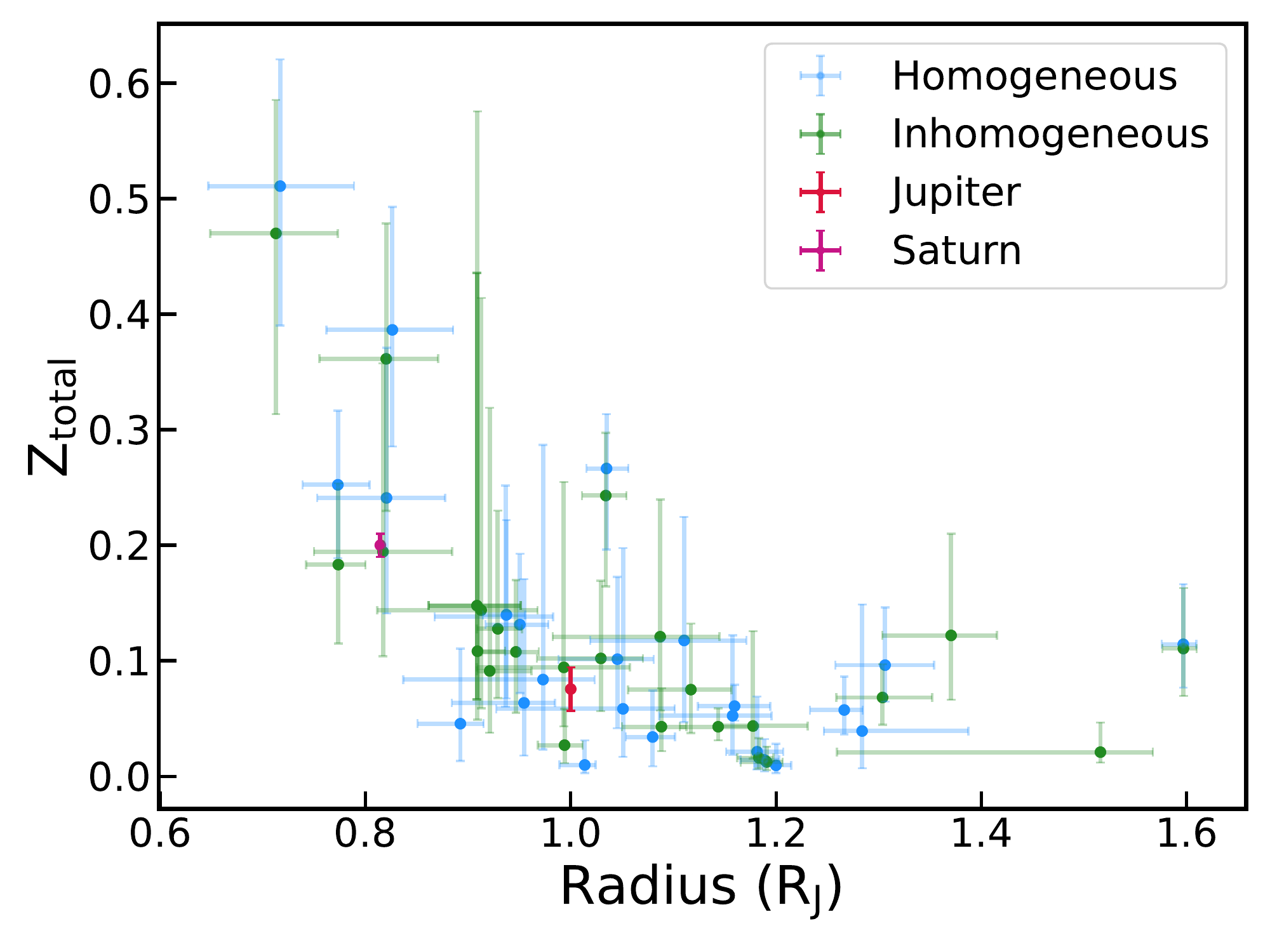}
    \caption{Bulk metallicity as a function of the best-fit radius. We see a clear correlation, with the heavy element fraction decreasing with the radius. The light blue points are the homogeneous runs and the dark green points are the inhomogeneous models. Jupiter and Saturn are plotted for reference \citep{miguel2022, Ni2020}. The error bars represent the 1$\sigma$ uncertainty.}
    \label{fig:core_rad}
\end{figure}

Our results show that there is a correlation between the core mass fraction and the radius of the planet. We show the best-fit core mass fraction as a function of the best-fit radius in Figure~\ref{fig:core_rad}, where we see that larger planets have smaller core mass fractions, a result that is also in agreement with trends found in the giant planets in the solar system \citep{miguel2022, Ni2020}.
In addition, another thing to notice is that smaller planets are less constrained than the larger planets in the sample. This is because the smaller planets have a larger amount of metals and larger degeneracies in the determination of their interior structures. 
Furthermore, the inhomogeneous models are often less constrained than the homogeneous models, which is caused by the difference in the size of the parameter space between the two models.

For most planets in the sample, we find that the core mass fraction is often very similar between runs of the same planet. For WASP-39\,b, for example, we show the found core mass fraction in Table~\ref{tab:wasp39b} for both the homogeneous and the inhomogeneous run. The values of both runs are very similar.
\begin{table}
\def\arraystretch{1.3}
	\centering
	\begin{tabular}{|l|rr}
	\hline 
	\hline 
	 & Homogeneous & Inhomogeneous \cr
	\hline 

Core mass fraction   & 0.0122 $^{+0.0158}_{-0.00867}$  & 0.0125 $^{+0.0142}_{-0.00861}$ \cr
Core mass ($M_{\textrm{earth}}$) & 0.88 $^{+1.1}_{-0.62}$  & 0.9 $^{+1}_{-0.61}$ \cr
	\hline 
	\hline 
	\end{tabular}
\caption{Core masses for both models fit to WASP-39\,b with 1$\sigma$ error bars.}
\label{tab:wasp39b}
\end{table}

\subsubsection{Mass metallicity relation for exoplanets}
\begin{figure}
    \centering
    \includegraphics[width=0.95\columnwidth]{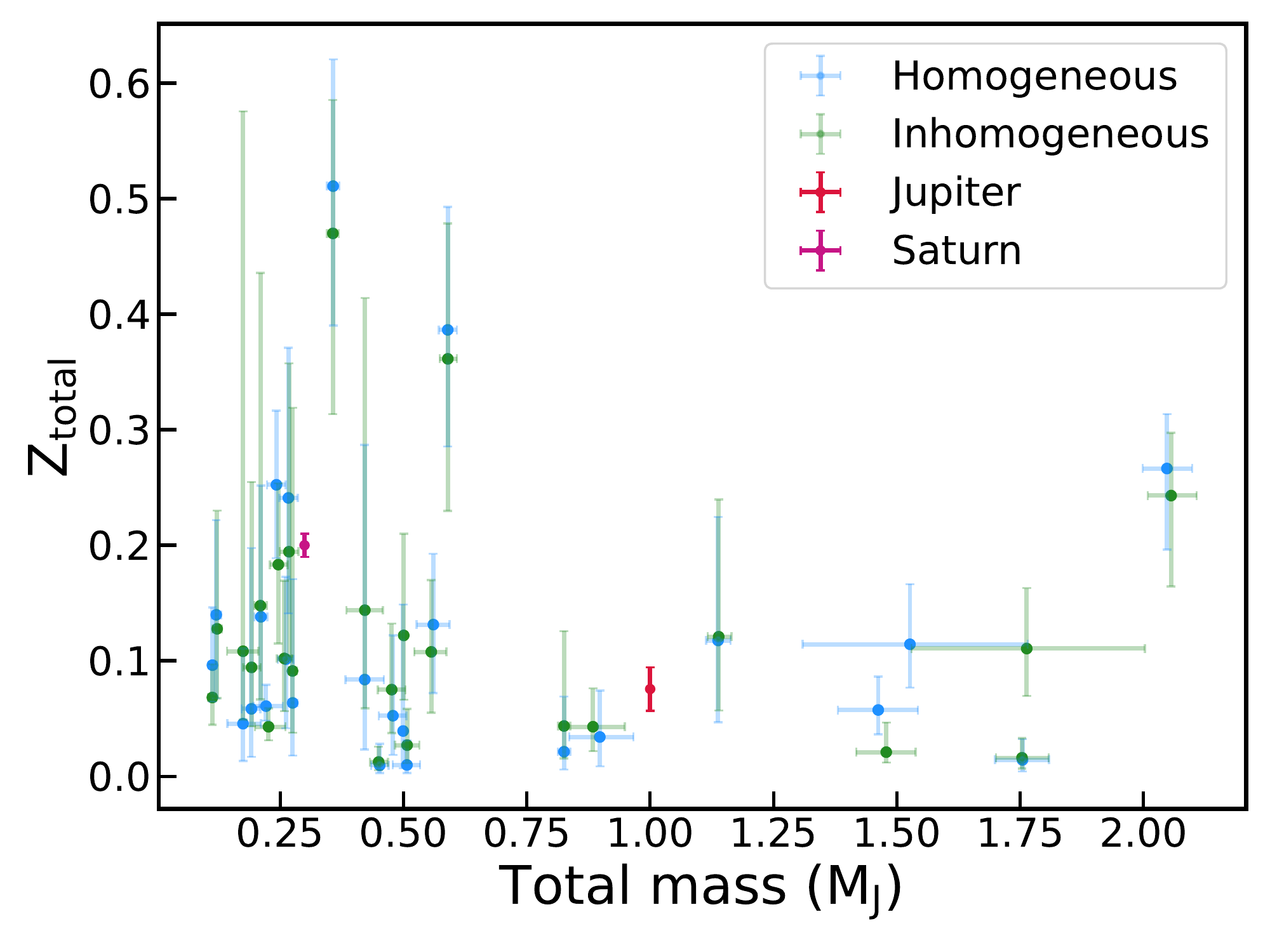}
    \caption{Bulk metallicities as a function of the total estimated planetary mass. The light blue points are the homogeneous runs and the dark green points are the inhomogeneous models. Jupiter and Saturn are shown in the figure for comparison. The error bars represent the 1$\sigma$ uncertainty.}
    \label{fig:mz_mass}
\end{figure}

We show the retrieved bulk metallicity of the exoplanets as a function of their mass in Figure~\ref{fig:mz_mass}. We notice that the metal mass fraction of a planet decreases as the total mass of the planet increases, a result in agreement with previous calculations \citep[e.g.][]{thorngren2016, thorngren2019}. We note that our retrieved metallicities are often slightly lower than the values provided by previous calculations (see section \ref{sec:comparison}). Our results are also in agreement with recent estimations of solar system giants \citep{miguel2022,mankovich2021}.

\begin{figure}
    \centering
    \includegraphics[width=0.95\columnwidth]{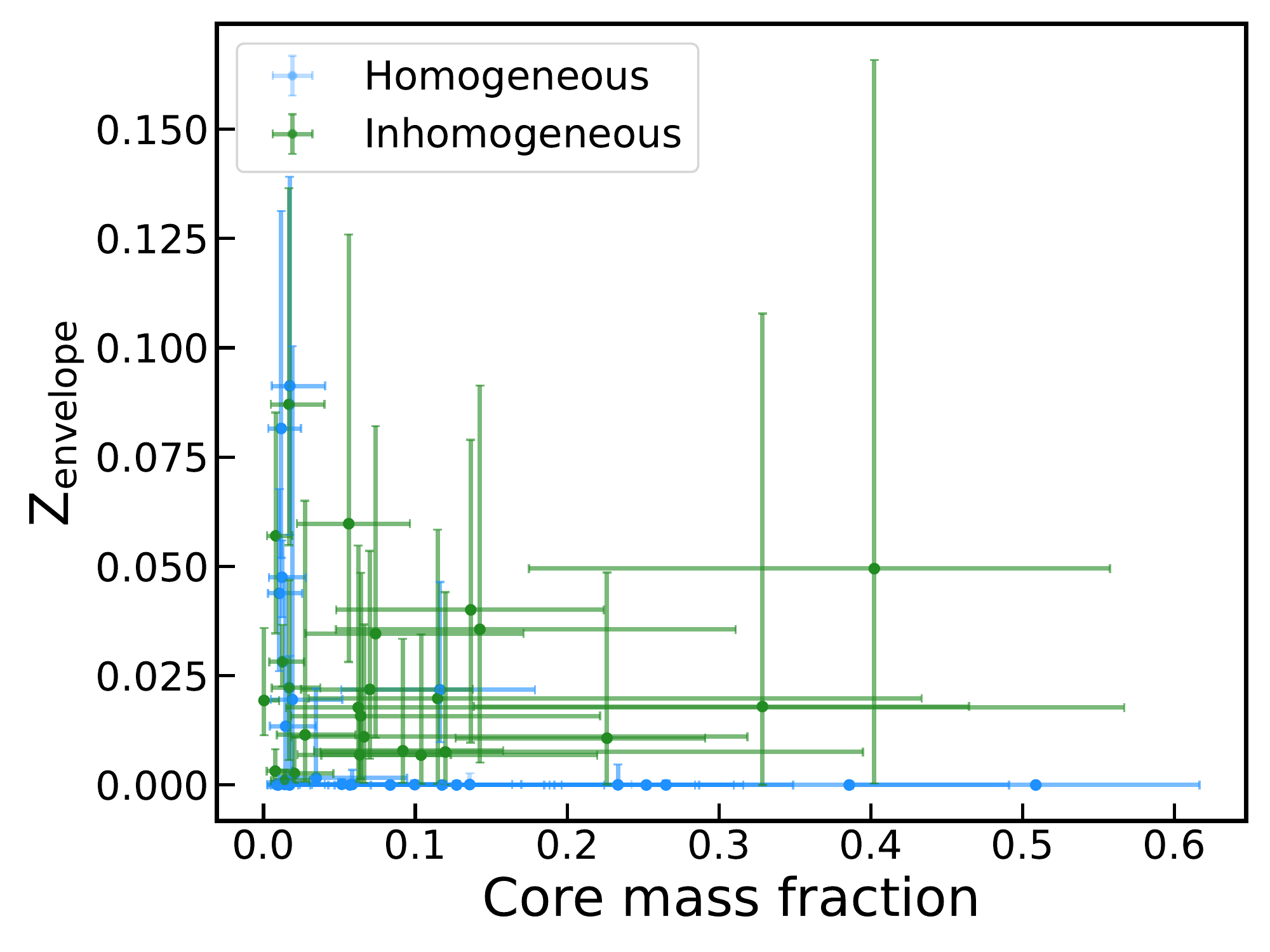}
    \caption{Distribution of metals over the core and envelope. The vertical axis shows the envelope heavy metal fraction and the horizontal axis shows the core mass fraction. The light blue points are the homogeneous runs and the dark green points are the inhomogeneous models. The error bars represent the 1$\sigma$ uncertainty.}
    \label{fig:mz_mcore}
\end{figure}

The inhomogeneous models could reproduce the homogeneous model solutions by finding solutions where the core-boundary metal fraction is equal to the atmospheric metal fraction. In that case, the distribution of the heavy elements over the core and the envelope should be equal to that in the homogeneous model. Figure~\ref{fig:mz_mcore} shows the metallicity in the envelope as a function of the core mass fraction, where we see that homogeneous and inhomogeneous models occupy different regions of the parameter space. The homogeneous solutions either have a relatively high core mass fraction and a low envelope metal mass fraction, or a low core mass fraction and a relatively high envelope metal mass fraction. The inhomogeneous models occupy the regions between these two extremes. This result implies that the inhomogeneous solutions are different from the homogeneous models, indicating that the homogeneous model is not the only possible solution to explain the interior structure of these planets.


\begin{figure}
    \centering
    \includegraphics[width=0.95\columnwidth]{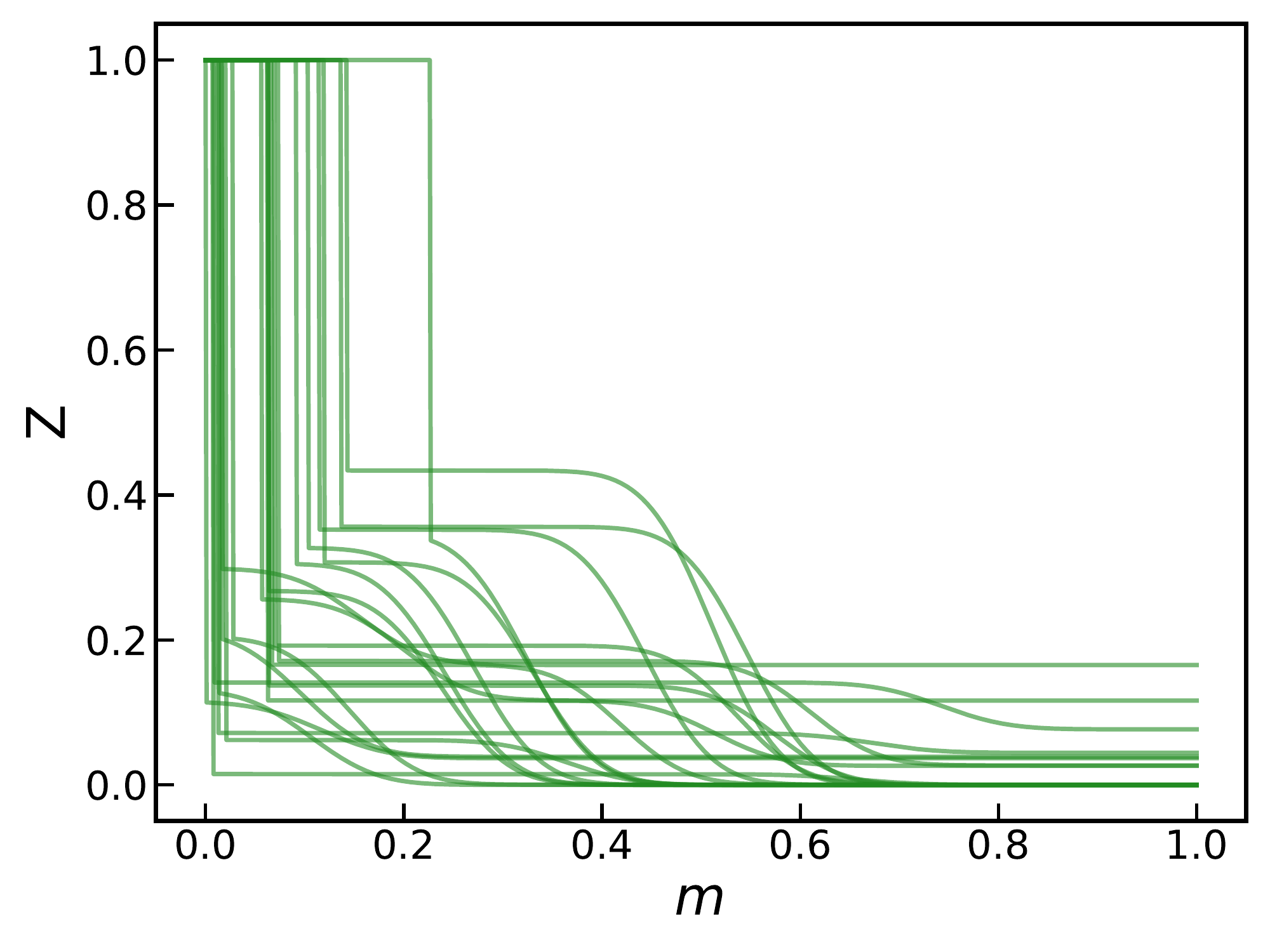}
    \caption{Heavy element fraction as a function of the mass coordinate of the inhomogeneous runs. The vertical axis shows the heavy element fraction and the horizontal axis shows the mass fraction. We only include the runs where the core mass fraction is below 0.25, to show the structure of the other planets more clearly.}
    \label{fig:dilute}
\end{figure}

To illustrate how the heavy elements are distributed in the envelope in inhomogeneous models, we plot the heavy element fraction as a function of mass of the best-fit models in Figure~\ref{fig:dilute}. We see that in most planets in our sample, the gradient of metals in the envelope extends in a range from 0.2 to 0.7 in mass fraction coordinates. Values found for solar system giants show that the gradient of metals or dilute core region (as this region is normally called in planetary science) extends between 0.5 and 0.6 of the planetary radius \citep{Wahl2016, miguel2022, mankovich2021}. On the other hand, formation models find that this region extends to a smaller radius, close to 0.3 radii \cite[e.g.][ and references therein]{helled2022,muller}. However, a recent analysis from \citet{Howard2023} points towards a better agreement between interior and formation models about the extent of the heavy-element gradient. The values found in this work for exoplanets cover the entire range of both theoretical and observational values, although these models do not have such tight constraints as the solar system giants.

\section{Discussion}
\label{sec:discuss}

\subsection{Comparison to previous work}\label{sec:comparison}
Previously, mass-metallicity relations for exoplanets have been estimated, for example by \citet{thorngren2016}. However, they use a different approach, considering evolution models (instead of static models as considered in this work), only using a homogeneous structure and putting at most 10\,M$_{\textrm{earth}}$ into the core. In general, our trend indicates slightly lower metallicities when looking at the entire sample of exoplanets, but we can also perform a direct comparison between our retrieved metal masses to their findings, as several exoplanets are included in both papers. An example is WASP-29\,b. For this planet, we find a metal mass of 19$^{+5.5}_{-5.1}$\,$M_{\textrm{earth}}$ using the homogeneous model and 14$^{+6.0}_{-5.4}$\,$M_{\textrm{earth}}$ with the inhomogeneous model. \citet{thorngren2016} find a metal mass of 25.65$^{+6.63}_{-6.09} M_{\textrm{earth}}$. The metal mass found in this work is lower, although the values agree within 3$\sigma$. The other planets included in both samples are HAT-P-12\,b, HAT-P-18\,b, WASP-69\,b and WASP-80\,b. We find the same metal mass for HAT-P-12\,b, close to the same value for HAT-P-18\,b and a higher value for WASP-69\,b and WASP-80\,b. All values agree within 3$\sigma$.

\citet{thorngren2019} also look at the total metal fraction of exoplanets. Their Figure~2 can be compared against our Figure~\ref{fig:mz_mass}. In both figures, Jupiter and Saturn are marked for reference. In the figure by \citet{thorngren2019}, Jupiter and Saturn are both significantly lower in metal fraction than the exoplanets in the sample. The points plotted for the exoplanets are limits, but the mean values are still higher than Jupiter and Saturn. In our figure, Jupiter and Saturn are in the middle of our sample and follow the trend established by the exoplanets. In our sample, there is no significant difference in the metal mass trend between the solar system and exoplanets.

There are several planets included in both samples which we can directly compare. These planets include WASP-39\,b, WASP-43\,b, WASP-52\,b and WASP-107\,b. We find lower metal mass fractions for all these planets, although the error bars are larger on our determined values.

\subsection{Homogeneous versus inhomogeneous structures}
A consequence of using the fitting method described in Section\,\ref{sec:fitting} is that it produces an evidence value for each run. Comparing these evidence values could give us information about which model describes the data better. The Bayesian evidence takes the size of the parameter space into account to reduce the risk of over-fitting. In practice, this means that two models with the same evidence value are equally likely, even if one model has more free parameters than the other.

\begin{figure}
    \centering
    \includegraphics[width=0.95\columnwidth]{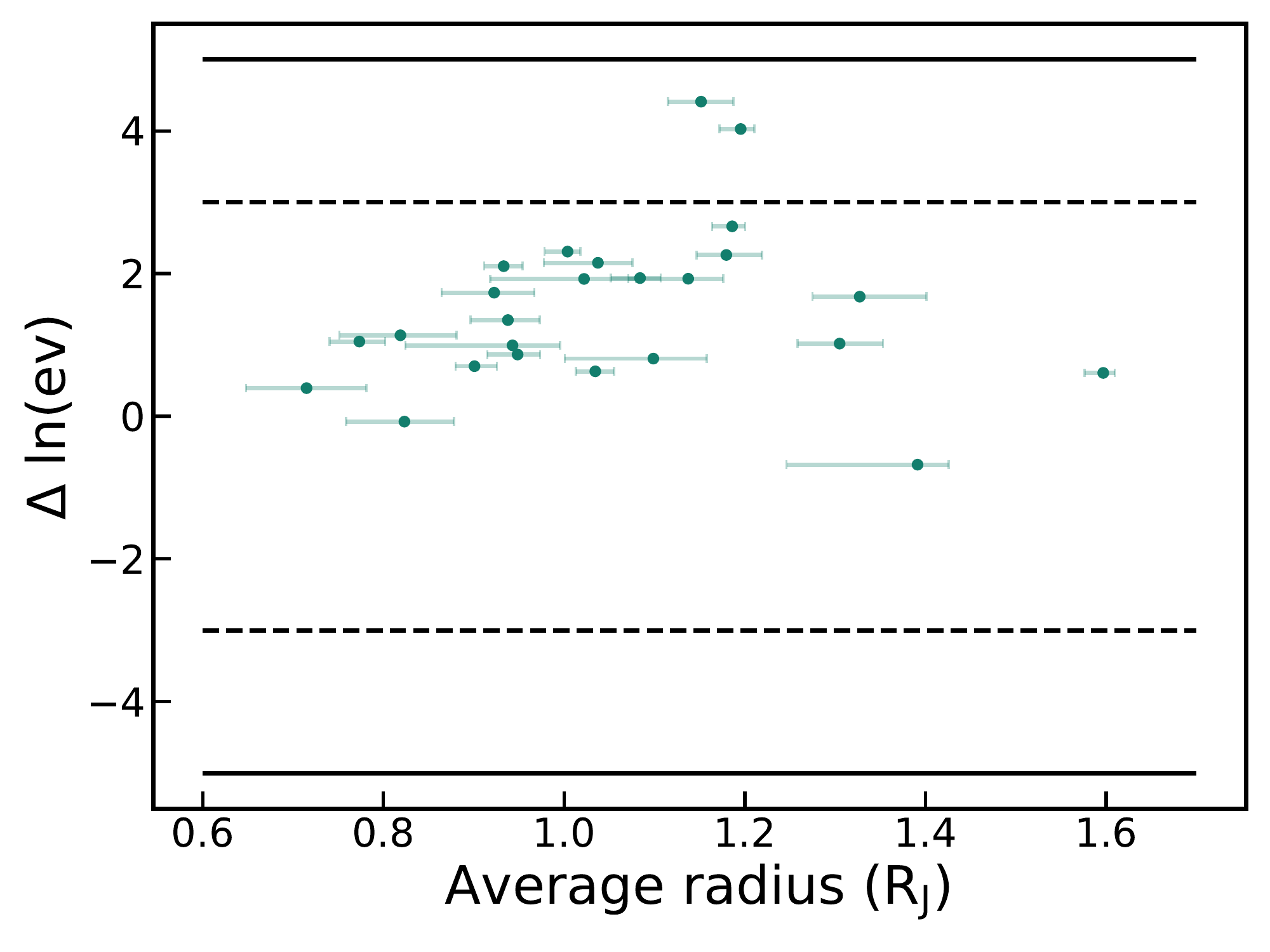}
    \caption{Difference in log evidence values between the homogeneous and inhomogeneous model runs. The horizontal axis is the average radius of the homogeneous and inhomogeneous runs. The solid black lines represent $\Delta$ln(Z$_{\textrm{ev}}$)=$\pm$5, and the dashed lines indicate $\Delta$ln(Z$_{\textrm{ev}}$)=$\pm$3. The error bars represent the 1$\sigma$ uncertainty.}
    \label{fig:evs}
\end{figure}

We list the found log evidence values for our simulated planet in Table~\ref{tab:evs_test_planet}. 
We see that there is no significant difference between the evidence values of the two models. This indicates that the parameter space is still too large for the fitting routine to find a unique solution. There are simply too many degeneracies to find one single solution. We see the same for our entire sample, as shown in Figure\,\ref{fig:evs}. Although some planets show a suggestion that one model is better than the other, none show strong evidence. More observational and theoretical constraints would be needed to determine which one of these two models fits exoplanet data better.

The observational constraints that could help include better measurements of the surface parameters of exoplanets, such as the metallicity, as well as measurements of Love numbers \citep[e.g.][]{lovenumbers}, which contain information on the interior structure.

Finally, the internal luminosity of giant exoplanets, especially those close to their host stars, is still not completely understood. Theoretical advancements on the range on this parameter would decrease the error bars on the parameters determined in this work.

\subsection{Impact of stellar metallicity}

\begin{figure}
    \centering
    \includegraphics[width=0.95\columnwidth]{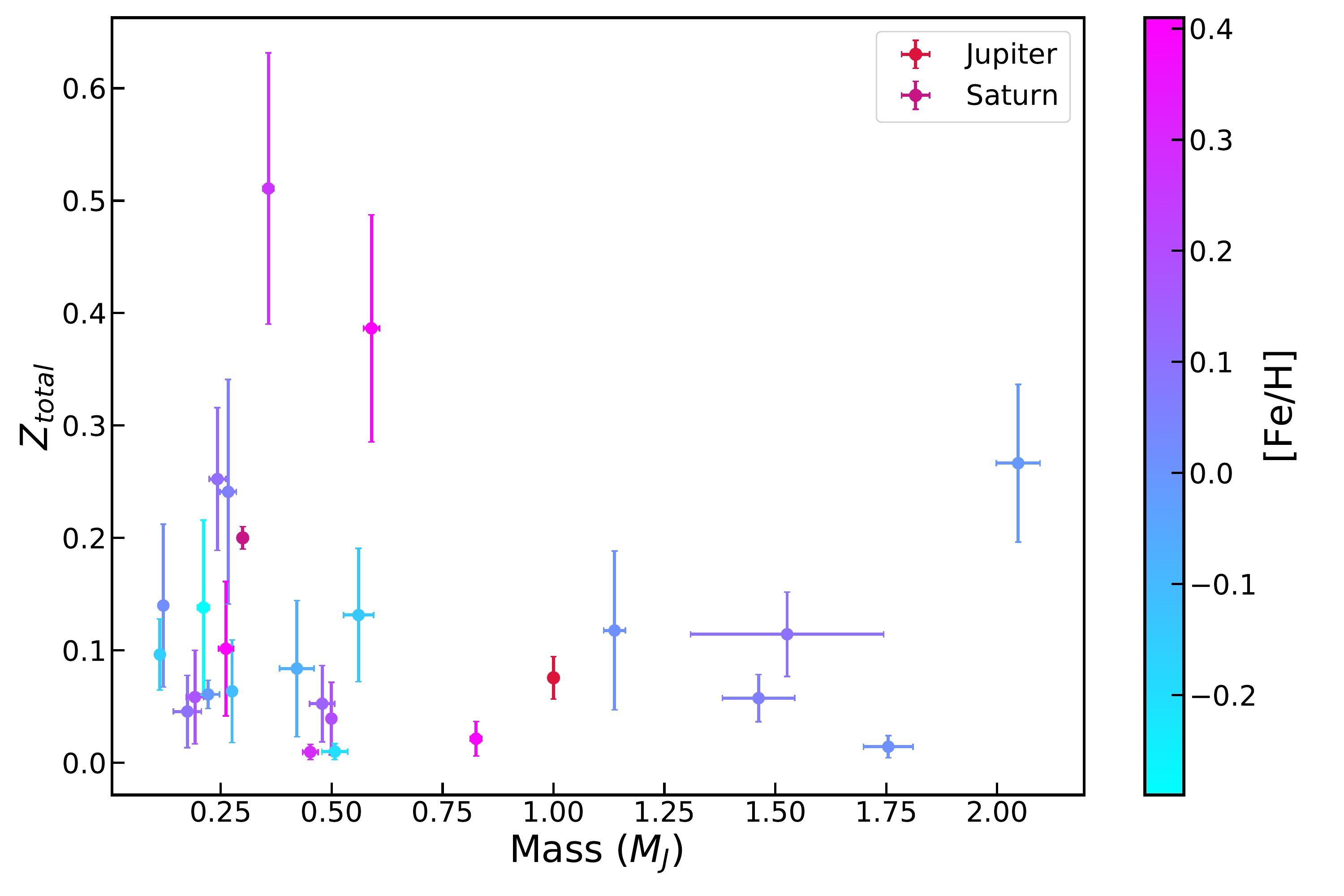}
    \caption{Bulk metallicities as a function of the total estimated planetary mass for the homogeneous runs, where the colour indicates the stellar metallicity. Jupiter and Saturn are shown in the figure for comparison. The error bars represent the 1$\sigma$ uncertainty. Metallicities from \citet{jwstmetals,2009ApJ...706..785H,2011ApJ...726...52H,2012ApJ...757..161T,2011AJ....141..179C,2012PASJ...64...97S,Pepper_2017,2014MNRAS.440.1982H,2014A&A...562L...3G,2017A&A...604A.110A,2014A&A...568A..81L,2010ApJ...723L..60H,10.1111/j.1365-2966.2010.16922.x,2012A&A...542A...4G,2009A&A...501..785G,2014A&A...568A.127M,2015AJ....150...18H,2013A&A...551A..80T,2014MNRAS.440.1982H}.}
    \label{fig:stellar}
\end{figure}

The metallicity of the star in a planetary system may affect the interior properties of the planets in that system. To test this hypothesis, we plot the total metal fraction as a function of mass, coloured by the metallicity of the star, in Fig~\ref{fig:stellar}. We show only the homogeneous model runs, but the results for the inhomogeneous model are almost identical.
Fig~\ref{fig:stellar} shows that the two planets with the highest metal mass fraction are also the two systems with the highest stellar metallicity. On average, it seems that planets orbiting a star with a higher metallicity than the Sun have a higher total metal fraction than the solar system planets, and the other way around. Although this correlation is not yet very significant, it suggests that the properties of the planet and its star are related. This will be further investigated in a future publication.

\subsection{Inflated planets}
\label{sec:inflated}

A significant amount of the exoplanets in our sample are inflated planets, with measured radii higher than our estimated values. To properly model these planets, we need to add more free parameters in our sample related to the atmospheric boundary condition and the energy deposited in the interior of the planet, which could increase the degeneracy in the system and the uncertainties on other parameters.
To test the effect of this, we re-ran all the inflated planets using a higher upper limit on the internal luminosity. Instead of limiting it to the values predicted by the evolution runs, we now set a general upper limit to the high (and arbitrary) value of $10^{29}$\,erg\,s$^{-1}$, or $3\cdot10^4 L_{\textrm{J}}$, for all the inflated planets.  
We see that with this new parameter space, we can reproduce the radius of even the most inflated planets very well. For comparison, we discuss two planets in more detail.

HAT-P-32\,b is one of the most inflated planets in our sample. It has a measured radius of 1.8\,$R_{\textrm{J}}$, but in the original runs, the best-fit radius is only 1\,$R_{\textrm{J}}$. When we increase the limit on the internal luminosity, the radius can get to 1.8\,$R_{\textrm{J}}$. However, now the core mass fraction and other free parameters are less constrained. The values of the core mass fraction and the radius are shown in Table~\ref{tab:hatp32b_high_lum} (for the homogeneous run). We have also added an intermediate run with the upper limit on the internal luminosity set to $10^{28}$\,erg\,s$^{-1}$, or $3\cdot10^3 L_{\textrm{J}}$. We see that this run still manages to reproduce the radius, and that the parameters are much more constrained than in the run with a higher internal luminosity limit.

\begin{table}
    \centering
    \def\arraystretch{1.5}
    \begin{tabular}{|l|rr|}
    \hline
    \hline
          & Radius ($R_{\textrm{J}}$)& Core mass fraction \cr
         \hline
         Default limit & 1.01$_{-0.19}^{+0.0548}$ & 0.144$^{+0.254}_{-0.0999}$  \cr
         $10^{29}$ erg\,s$^{-1}$ limit & 1.79$_{-0.0306}^{+0.0255}$ & 0.194$^{+0.0975}_{-0.0914}$\cr
         $10^{28}$ erg\,s$^{-1}$ limit & 1.78$_{-0.0322}^{+0.0239}$ & 0.0428$^{+0.0693}_{-0.0302}$ \cr
    \hline
    \end{tabular}
    \caption{Comparison of different internal luminosity upper limits for HAT-P-32\,b, for the homogeneous runs. The given error bars are 1$\sigma$.}
    \label{tab:hatp32b_high_lum}
\end{table}

HD\,209458\,b is a well-known inflated planet. We find that again, for this planet, we cannot reproduce the observed radius using our method. However, with the increased internal luminosity limit at $10^{29}$\,erg\,s$^{-1}$, we can reproduce the radius, as shown in Table~\ref{tab:hd209458b_high_lum}. 
When we decrease the limit to $10^{28}$\,erg\,s$^{-1}$, the parameters are slightly more constrained.

\begin{table}
    \centering
    \def\arraystretch{1.5}
    \begin{tabular}{|l|rr|}
    \hline
    \hline
          & Radius ($R_{\textrm{J}}$)& Core mass fraction \cr
         \hline
         Default limit & 1.05$_{-0.0525}^{+0.0197}$ & 0.286$^{+0.0603}_{-0.0208}$  \cr
         $10^{29}$ erg\,s$^{-1}$ limit & 1.38$_{-0.0195}^{+0.0192}$ & 0.390$^{+0.0591}_{-0.0913}$\cr
         $10^{28}$ erg\,s$^{-1}$ limit & 1.38$_{-0.0216}^{+0.0187}$ & 0.155$^{+0.0433}_{-0.0756}$ \cr
    \hline
    \hline
    \end{tabular}
    \caption{Comparison of different internal luminosity upper limits for HD\,209458\,b, for the homogeneous runs. The given error bars are 1$\sigma$.}
    \label{tab:hd209458b_high_lum}
\end{table}

The two values used here for the upper limits are arbitrary values chosen to test the impact of increasing the internal luminosity. However, what this test shows is that any inflation mechanism that relies on adding a free parameter to the model will increase the parameter space and make it more difficult to constrain the interior structures of exoplanets. If this mechanism is understood well enough and constrained based on the properties of the planet, it would most likely be possible to recover the interior structure.

\section{Conclusions}
\label{sec:concl}
In this work, we present a method based on static calculations to retrieve the interior properties of exoplanets, using observational data as constraints, including their masses, radii, equilibrium temperatures and atmospheric metallicities. In our model, we adopt the traditionally used homogeneous structures of an exoplanet, where the interior consists of a core and a homogeneous envelope \citep[e.g.][]{thorngren2016, thorngren2019}, but also, based on recent results from the giant planets in the solar system \citep{mankovich2021, miguel2022} and on formation models of giant planets \citep{lozovsky2018, valletta2022}, we introduce inhomogeneous structures to the study of exoplanet interiors. In these inhomogeneous models, the metals in the envelope decrease gradually from the core toward the atmosphere, creating more diverse interior structures with a different distribution of metals and bulk metallicities.

We apply the method to a sample of exoplanets and retrieve their core mass fractions, envelope metal fractions and bulk metallicities. Our results show that for larger planets, we find smaller core mass fractions. An analysis of the mass-metallicity relation for both models (homogeneous and inhomogeneous structures) shows that our sample of planets is in excellent agreement with recent determination of bulk metallicities of the giant planets in the solar system \citep{mankovich2021, miguel2022}. We also see that more massive planets have a lower metal mass fraction, in agreement with previous estimations \citep{thorngren2016, thorngren2019}. 

Our results on a test planet, as well as on the entire sample of exoplanets, show that the method provides accurate constraints for exoplanets that are not inflated, but that extra constraints would be needed to use it in retrieval estimations of highly inflated planets. Both models presented here can describe the sample of exoplanets well. To distinguish between the two, we require more observational constraints, such as improved measurements of atmospheric parameters, as well as potentially Love numbers.

This study is important in the context of future JWST observations, where atmospheric metallicities would be found for more exoplanets and with more accurate estimations, which would help to constrain their interior properties and to inform formation models towards a better determination of the history of planetary systems.

\section*{Acknowledgements}
This research made use of NASA's Astrophysics Data System, the \textsc{IPython} package \citep{PER-GRA:2007}; \textsc{SciPy} \citep{scipy}; \textsc{Matplotlib}, a \textsc{Python} library for publication quality graphics \citep{Hunter:2007}; \textsc{Astropy}, a community-developed core \textsc{Python} package for astronomy \citep{2013A&A...558A..33A}; \textsc{corner}, a \textsc{Python} library for making corner plots; and \textsc{NumPy} \citep{van2011numpy}. 
This work was performed using the compute resources from the Academic Leiden Interdisciplinary Cluster Environment (ALICE) provided by Leiden University.

\section*{Data availability}
All best-fit results are available at \url{https://github.com/AstroYamila-Team/exoplanet-interior-retrievals}.

\bibliographystyle{mnras}
\bibliography{mybib_stars_paper.bib}

\begin{thebibliography}{}
\makeatletter
\relax
\def\mn@urlcharsother{\let\do\@makeother \do\$\do\&\do\#\do\^\do\_\do\%\do\~}
\def\mn@doi{\begingroup\mn@urlcharsother \@ifnextchar [ {\mn@doi@}
  {\mn@doi@[]}}
\def\mn@doi@[#1]#2{\def\@tempa{#1}\ifx\@tempa\@empty \href
  {http://dx.doi.org/#2} {doi:#2}\else \href {http://dx.doi.org/#2} {#1}\fi
  \endgroup}
\def\mn@eprint#1#2{\mn@eprint@#1:#2::\@nil}
\def\mn@eprint@arXiv#1{\href {http://arxiv.org/abs/#1} {{\tt arXiv:#1}}}
\def\mn@eprint@dblp#1{\href {http://dblp.uni-trier.de/rec/bibtex/#1.xml}
  {dblp:#1}}
\def\mn@eprint@#1:#2:#3:#4\@nil{\def\@tempa {#1}\def\@tempb {#2}\def\@tempc
  {#3}\ifx \@tempc \@empty \let \@tempc \@tempb \let \@tempb \@tempa \fi \ifx
  \@tempb \@empty \def\@tempb {arXiv}\fi \@ifundefined
  {mn@eprint@\@tempb}{\@tempb:\@tempc}{\expandafter \expandafter \csname
  mn@eprint@\@tempb\endcsname \expandafter{\@tempc}}}

\bibitem[\protect\citeauthoryear{{Anderson} et~al.,}{{Anderson}
  et~al.}{2010}]{2010ApJ...709..159A}
{Anderson} D.~R.,  et~al., 2010, \mn@doi [\apj] {10.1088/0004-637X/709/1/159},
  \href {https://ui.adsabs.harvard.edu/abs/2010ApJ...709..159A} {709, 159}

\bibitem[\protect\citeauthoryear{{Anderson} et~al.,}{{Anderson}
  et~al.}{2011a}]{2011MNRAS.416.2108A}
{Anderson} D.~R.,  et~al., 2011a, \mn@doi [\mnras]
  {10.1111/j.1365-2966.2011.19182.x}, \href
  {https://ui.adsabs.harvard.edu/abs/2011MNRAS.416.2108A} {416, 2108}

\bibitem[\protect\citeauthoryear{{Anderson} et~al.,}{{Anderson}
  et~al.}{2011b}]{2011A&A...531A..60A}
{Anderson} D.~R.,  et~al., 2011b, \mn@doi [\aap] {10.1051/0004-6361/201016208},
  \href {https://ui.adsabs.harvard.edu/abs/2011A&A...531A..60A} {531, A60}

\bibitem[\protect\citeauthoryear{{Anderson} et~al.,}{{Anderson}
  et~al.}{2014}]{2014MNRAS.445.1114A}
{Anderson} D.~R.,  et~al., 2014, \mn@doi [\mnras] {10.1093/mnras/stu1737},
  \href {https://ui.adsabs.harvard.edu/abs/2014MNRAS.445.1114A} {445, 1114}

\bibitem[\protect\citeauthoryear{{Anderson} et~al.,}{{Anderson}
  et~al.}{2017}]{2017A&A...604A.110A}
{Anderson} D.~R.,  et~al., 2017, \mn@doi [\aap] {10.1051/0004-6361/201730439},
  \href {https://ui.adsabs.harvard.edu/abs/2017A&A...604A.110A} {604, A110}

\bibitem[\protect\citeauthoryear{Anisman, Edwards, Changeat, Venot, Al-Refaie,
  Tsiaras  \& Tinetti}{Anisman et~al.}{2020}]{Anisman_2020}
Anisman L.~O.,  Edwards B.,  Changeat Q.,  Venot O.,  Al-Refaie A.~F.,  Tsiaras
  A.,   Tinetti G.,  2020, \mn@doi [The Astronomical Journal]
  {10.3847/1538-3881/abb9b0}, 160, 233

\bibitem[\protect\citeauthoryear{{Astropy Collaboration} et~al.,}{{Astropy
  Collaboration} et~al.}{2013}]{2013A&A...558A..33A}
{Astropy Collaboration} et~al., 2013, \mn@doi [\aap]
  {10.1051/0004-6361/201322068}, \href
  {http://adsabs.harvard.edu/abs/2013A%26A...558A..33A} {558, A33}

\bibitem[\protect\citeauthoryear{{Bakos} et~al.,}{{Bakos}
  et~al.}{2007}]{2007ApJ...656..552B}
{Bakos} G.~{\'A}.,  et~al., 2007, \mn@doi [\apj] {10.1086/509874}, \href
  {https://ui.adsabs.harvard.edu/abs/2007ApJ...656..552B} {656, 552}

\bibitem[\protect\citeauthoryear{{Birkby}, {de Kok}, {Brogi}, {Schwarz}  \&
  {Snellen}}{{Birkby} et~al.}{2017}]{2017AJ....153..138B}
{Birkby} J.~L.,  {de Kok} R.~J.,  {Brogi} M.,  {Schwarz} H.,   {Snellen}
  I.~A.~G.,  2017, \mn@doi [\aj] {10.3847/1538-3881/aa5c87}, \href
  {https://ui.adsabs.harvard.edu/abs/2017AJ....153..138B} {153, 138}

\bibitem[\protect\citeauthoryear{{Bloot}, {Callingham}  \& {Marcote}}{{Bloot}
  et~al.}{2022}]{Bloot2022}
{Bloot} S.,  {Callingham} J.~R.,   {Marcote} B.,  2022, \mn@doi [\mnras]
  {10.1093/mnras/stab2976}, \href
  {https://ui.adsabs.harvard.edu/abs/2022MNRAS.509..475B} {509, 475}

\bibitem[\protect\citeauthoryear{{Bonomo} et~al.,}{{Bonomo}
  et~al.}{2017}]{2017A&A...602A.107B}
{Bonomo} A.~S.,  et~al., 2017, \mn@doi [\aap] {10.1051/0004-6361/201629882},
  \href {https://ui.adsabs.harvard.edu/abs/2017A&A...602A.107B} {602, A107}

\bibitem[\protect\citeauthoryear{{Brogi}, {Snellen}, {de Kok}, {Albrecht},
  {Birkby}  \& {de Mooij}}{{Brogi} et~al.}{2013}]{2013ApJ...767...27B}
{Brogi} M.,  {Snellen} I.~A.~G.,  {de Kok} R.~J.,  {Albrecht} S.,  {Birkby}
  J.~L.,   {de Mooij} E.~J.~W.,  2013, \mn@doi [\apj]
  {10.1088/0004-637X/767/1/27}, \href
  {https://ui.adsabs.harvard.edu/abs/2013ApJ...767...27B} {767, 27}

\bibitem[\protect\citeauthoryear{Bruno et~al.,}{Bruno
  et~al.}{2018}]{Bruno_2018}
Bruno G.,  et~al., 2018, \mn@doi [The Astronomical Journal]
  {10.3847/1538-3881/aaa0c7}, 155, 55

\bibitem[\protect\citeauthoryear{{Buchner} et~al.,}{{Buchner}
  et~al.}{2014}]{2014A&A...564A.125B}
{Buchner} J.,  et~al., 2014, \mn@doi [\aap] {10.1051/0004-6361/201322971},
  \href {https://ui.adsabs.harvard.edu/abs/2014A&A...564A.125B} {564, A125}

\bibitem[\protect\citeauthoryear{{Callingham} et~al.,}{{Callingham}
  et~al.}{2015}]{Callingham2015}
{Callingham} J.~R.,  et~al., 2015, \mn@doi [\apj]
  {10.1088/0004-637X/809/2/168}, \href
  {http://adsabs.harvard.edu/abs/2015ApJ...809..168C} {809, 168}

\bibitem[\protect\citeauthoryear{Cameron et~al.,}{Cameron
  et~al.}{2010}]{10.1111/j.1365-2966.2010.16922.x}
Cameron A.~C.,  et~al., 2010, \mn@doi [Monthly Notices of the Royal
  Astronomical Society] {10.1111/j.1365-2966.2010.16922.x}, 407, 507

\bibitem[\protect\citeauthoryear{Chakrabarty \& Sengupta}{Chakrabarty \&
  Sengupta}{2019}]{Chakrabarty_2019}
Chakrabarty A.,  Sengupta S.,  2019, \mn@doi [The Astronomical Journal]
  {10.3847/1538-3881/ab24dd}, 158, 39

\bibitem[\protect\citeauthoryear{{Chan}, {Ingemyr}, {Winn}, {Holman},
  {Sanchis-Ojeda}, {Esquerdo}  \& {Everett}}{{Chan}
  et~al.}{2011}]{2011AJ....141..179C}
{Chan} T.,  {Ingemyr} M.,  {Winn} J.~N.,  {Holman} M.~J.,  {Sanchis-Ojeda} R.,
  {Esquerdo} G.,   {Everett} M.,  2011, \mn@doi [\aj]
  {10.1088/0004-6256/141/6/179}, \href
  {https://ui.adsabs.harvard.edu/abs/2011AJ....141..179C} {141, 179}

\bibitem[\protect\citeauthoryear{{Chandrasekhar}}{{Chandrasekhar}}{1935}]{Chandrasekhar1935}
{Chandrasekhar} S.,  1935, \mn@doi [\mnras] {10.1093/mnras/96.1.21}, \href
  {https://ui.adsabs.harvard.edu/abs/1935MNRAS..96...21C} {96, 21}

\bibitem[\protect\citeauthoryear{{Chen} et~al.,}{{Chen}
  et~al.}{2018}]{2018A&A...616A.145C}
{Chen} G.,  et~al., 2018, \mn@doi [\aap] {10.1051/0004-6361/201833033}, \href
  {https://ui.adsabs.harvard.edu/abs/2018A&A...616A.145C} {616, A145}

\bibitem[\protect\citeauthoryear{{Collins}, {Kielkopf}  \& {Stassun}}{{Collins}
  et~al.}{2017}]{2017AJ....153...78C}
{Collins} K.~A.,  {Kielkopf} J.~F.,   {Stassun} K.~G.,  2017, \mn@doi [\aj]
  {10.3847/1538-3881/153/2/78}, \href
  {https://ui.adsabs.harvard.edu/abs/2017AJ....153...78C} {153, 78}

\bibitem[\protect\citeauthoryear{Col{\'{o}}n et~al.,}{Col{\'{o}}n
  et~al.}{2020}]{Col_n_2020}
Col{\'{o}}n K.~D.,  et~al., 2020, \mn@doi [The Astronomical Journal]
  {10.3847/1538-3881/abc1e9}, 160, 280

\bibitem[\protect\citeauthoryear{{Cort{\'e}s-Zuleta}, {Rojo}, {Wang}, {Hinse},
  {Hoyer}, {Sanhueza}, {Correa-Amaro}  \& {Albornoz}}{{Cort{\'e}s-Zuleta}
  et~al.}{2020}]{2020A&A...636A..98C}
{Cort{\'e}s-Zuleta} P.,  {Rojo} P.,  {Wang} S.,  {Hinse} T.~C.,  {Hoyer} S.,
  {Sanhueza} B.,  {Correa-Amaro} P.,   {Albornoz} J.,  2020, \mn@doi [\aap]
  {10.1051/0004-6361/201936279}, \href
  {https://ui.adsabs.harvard.edu/abs/2020A&A...636A..98C} {636, A98}

\bibitem[\protect\citeauthoryear{{Delrez} et~al.,}{{Delrez}
  et~al.}{2016}]{2016MNRAS.458.4025D}
{Delrez} L.,  et~al., 2016, \mn@doi [\mnras] {10.1093/mnras/stw522}, \href
  {https://ui.adsabs.harvard.edu/abs/2016MNRAS.458.4025D} {458, 4025}

\bibitem[\protect\citeauthoryear{{Esposito, M.} et~al.,}{{Esposito, M.}
  et~al.}{2014}]{refId0}
{Esposito, M.} et~al., 2014, \mn@doi [A\&A] {10.1051/0004-6361/201423735}, 564,
  L13

\bibitem[\protect\citeauthoryear{{Faedi} et~al.,}{{Faedi}
  et~al.}{2011}]{2011A&A...531A..40F}
{Faedi} F.,  et~al., 2011, \mn@doi [\aap] {10.1051/0004-6361/201116671}, \href
  {https://ui.adsabs.harvard.edu/abs/2011A&A...531A..40F} {531, A40}

\bibitem[\protect\citeauthoryear{{Feroz}, {Hobson}, {Cameron}  \&
  {Pettitt}}{{Feroz} et~al.}{2013}]{Feroz2013}
{Feroz} F.,  {Hobson} M.~P.,  {Cameron} E.,   {Pettitt} A.~N.,  2013,
  arXiv:1306.2144, \href {http://adsabs.harvard.edu/abs/2013arXiv1306.2144F} {}

\bibitem[\protect\citeauthoryear{{Gibson}, {Aigrain}, {Barstow}, {Evans},
  {Fletcher}  \& {Irwin}}{{Gibson} et~al.}{2013}]{2013MNRAS.428.3680G}
{Gibson} N.~P.,  {Aigrain} S.,  {Barstow} J.~K.,  {Evans} T.~M.,  {Fletcher}
  L.~N.,   {Irwin} P.~G.~J.,  2013, \mn@doi [\mnras] {10.1093/mnras/sts307},
  \href {https://ui.adsabs.harvard.edu/abs/2013MNRAS.428.3680G} {428, 3680}

\bibitem[\protect\citeauthoryear{{Gillon} et~al.,}{{Gillon}
  et~al.}{2009}]{2009A&A...501..785G}
{Gillon} M.,  et~al., 2009, \mn@doi [\aap] {10.1051/0004-6361/200911749}, \href
  {https://ui.adsabs.harvard.edu/abs/2009A&A...501..785G} {501, 785}

\bibitem[\protect\citeauthoryear{{Gillon} et~al.,}{{Gillon}
  et~al.}{2012}]{2012A&A...542A...4G}
{Gillon} M.,  et~al., 2012, \mn@doi [\aap] {10.1051/0004-6361/201218817}, \href
  {https://ui.adsabs.harvard.edu/abs/2012A&A...542A...4G} {542, A4}

\bibitem[\protect\citeauthoryear{{Gillon} et~al.,}{{Gillon}
  et~al.}{2014}]{2014A&A...562L...3G}
{Gillon} M.,  et~al., 2014, \mn@doi [\aap] {10.1051/0004-6361/201323014}, \href
  {https://ui.adsabs.harvard.edu/abs/2014A&A...562L...3G} {562, L3}

\bibitem[\protect\citeauthoryear{{Goyal}, {Lewis}, {Wakeford}, {MacDonald}  \&
  {Mayne}}{{Goyal} et~al.}{2021}]{2021ApJ...923..242G}
{Goyal} J.~M.,  {Lewis} N.~K.,  {Wakeford} H.~R.,  {MacDonald} R.~J.,   {Mayne}
  N.~J.,  2021, \mn@doi [\apj] {10.3847/1538-4357/ac27b2}, \href
  {https://ui.adsabs.harvard.edu/abs/2021ApJ...923..242G} {923, 242}

\bibitem[\protect\citeauthoryear{{Guillot} \& {Gautier}}{{Guillot} \&
  {Gautier}}{2015}]{guillot_gautier_2014}
{Guillot} T.,  {Gautier} D.,  2015, in {Schubert} G.,  ed., , Treatise on
  Geophysics.
pp 529--557, \mn@doi{10.1016/B978-0-444-53802-4.00176-7}

\bibitem[\protect\citeauthoryear{{Guillot} \& {Morel}}{{Guillot} \&
  {Morel}}{1995}]{guillot1995}
{Guillot} T.,  {Morel} P.,  1995, \aaps, \href
  {https://ui.adsabs.harvard.edu/abs/1995A&AS..109..109G} {109, 109}

\bibitem[\protect\citeauthoryear{{Guillot} \& {Showman}}{{Guillot} \&
  {Showman}}{2002}]{guillot2002}
{Guillot} T.,  {Showman} A.~P.,  2002, \mn@doi [\aap]
  {10.1051/0004-6361:20011624}, \href
  {https://ui.adsabs.harvard.edu/abs/2002A&A...385..156G} {385, 156}

\bibitem[\protect\citeauthoryear{{Guillot} et~al.,}{{Guillot}
  et~al.}{2018}]{Guillot2018}
{Guillot} T.,  et~al., 2018, \mn@doi [\nat] {10.1038/nature25775}, \href
  {https://ui.adsabs.harvard.edu/abs/2018Natur.555..227G} {555, 227}

\bibitem[\protect\citeauthoryear{{Hartman} et~al.,}{{Hartman}
  et~al.}{2009}]{2009ApJ...706..785H}
{Hartman} J.~D.,  et~al., 2009, \mn@doi [\apj] {10.1088/0004-637X/706/1/785},
  \href {https://ui.adsabs.harvard.edu/abs/2009ApJ...706..785H} {706, 785}

\bibitem[\protect\citeauthoryear{{Hartman} et~al.,}{{Hartman}
  et~al.}{2011a}]{2011ApJ...726...52H}
{Hartman} J.~D.,  et~al., 2011a, \mn@doi [\apj] {10.1088/0004-637X/726/1/52},
  \href {https://ui.adsabs.harvard.edu/abs/2011ApJ...726...52H} {726, 52}

\bibitem[\protect\citeauthoryear{{Hartman} et~al.,}{{Hartman}
  et~al.}{2011b}]{2011ApJ...742...59H}
{Hartman} J.~D.,  et~al., 2011b, \mn@doi [\apj] {10.1088/0004-637X/742/1/59},
  \href {https://ui.adsabs.harvard.edu/abs/2011ApJ...742...59H} {742, 59}

\bibitem[\protect\citeauthoryear{{Hartman} et~al.,}{{Hartman}
  et~al.}{2012}]{2012AJ....144..139H}
{Hartman} J.~D.,  et~al., 2012, \mn@doi [\aj] {10.1088/0004-6256/144/5/139},
  \href {https://ui.adsabs.harvard.edu/abs/2012AJ....144..139H} {144, 139}

\bibitem[\protect\citeauthoryear{{Hebb} et~al.,}{{Hebb}
  et~al.}{2010}]{2010ApJ...708..224H}
{Hebb} L.,  et~al., 2010, \mn@doi [\apj] {10.1088/0004-637X/708/1/224}, \href
  {https://ui.adsabs.harvard.edu/abs/2010ApJ...708..224H} {708, 224}

\bibitem[\protect\citeauthoryear{{H{\'e}brard} et~al.,}{{H{\'e}brard}
  et~al.}{2013}]{2013A&A...549A.134H}
{H{\'e}brard} G.,  et~al., 2013, \mn@doi [\aap] {10.1051/0004-6361/201220363},
  \href {https://ui.adsabs.harvard.edu/abs/2013A&A...549A.134H} {549, A134}

\bibitem[\protect\citeauthoryear{{Helled} et~al.,}{{Helled}
  et~al.}{2022}]{helled2022}
{Helled} R.,  et~al., 2022, \mn@doi [\icarus] {10.1016/j.icarus.2022.114937},
  \href {https://ui.adsabs.harvard.edu/abs/2022Icar..37814937H} {378, 114937}

\bibitem[\protect\citeauthoryear{{Hellier} et~al.,}{{Hellier}
  et~al.}{2010}]{2010ApJ...723L..60H}
{Hellier} C.,  et~al., 2010, \mn@doi [\apjl] {10.1088/2041-8205/723/1/L60},
  \href {https://ui.adsabs.harvard.edu/abs/2010ApJ...723L..60H} {723, L60}

\bibitem[\protect\citeauthoryear{{Hellier} et~al.,}{{Hellier}
  et~al.}{2012}]{2012MNRAS.426..739H}
{Hellier} C.,  et~al., 2012, \mn@doi [\mnras]
  {10.1111/j.1365-2966.2012.21780.x}, \href
  {https://ui.adsabs.harvard.edu/abs/2012MNRAS.426..739H} {426, 739}

\bibitem[\protect\citeauthoryear{{Hellier} et~al.,}{{Hellier}
  et~al.}{2014}]{2014MNRAS.440.1982H}
{Hellier} C.,  et~al., 2014, \mn@doi [\mnras] {10.1093/mnras/stu410}, \href
  {https://ui.adsabs.harvard.edu/abs/2014MNRAS.440.1982H} {440, 1982}

\bibitem[\protect\citeauthoryear{{Hellier} et~al.,}{{Hellier}
  et~al.}{2015}]{2015AJ....150...18H}
{Hellier} C.,  et~al., 2015, \mn@doi [\aj] {10.1088/0004-6256/150/1/18}, \href
  {https://ui.adsabs.harvard.edu/abs/2015AJ....150...18H} {150, 18}

\bibitem[\protect\citeauthoryear{{Holman} et~al.,}{{Holman}
  et~al.}{2006}]{2006ApJ...652.1715H}
{Holman} M.~J.,  et~al., 2006, \mn@doi [\apj] {10.1086/508155}, \href
  {https://ui.adsabs.harvard.edu/abs/2006ApJ...652.1715H} {652, 1715}

\bibitem[\protect\citeauthoryear{{Howard} et~al.,}{{Howard}
  et~al.}{2023}]{Howard2023}
{Howard} S.,  et~al., 2023, \mn@doi [\aap] {10.1051/0004-6361/202245625}, \href
  {https://ui.adsabs.harvard.edu/abs/2023A&A...672A..33H} {672, A33}

\bibitem[\protect\citeauthoryear{{Hubbard} \& {Marley}}{{Hubbard} \&
  {Marley}}{1989}]{Hubbard1989}
{Hubbard} W.~B.,  {Marley} M.~S.,  1989, \mn@doi [\icarus]
  {10.1016/0019-1035(89)90072-9}, \href
  {https://ui.adsabs.harvard.edu/abs/1989Icar...78..102H} {78, 102}

\bibitem[\protect\citeauthoryear{Hunter}{Hunter}{2007}]{Hunter:2007}
Hunter J.~D.,  2007, Computing In Science \& Engineering, 9, 90

\bibitem[\protect\citeauthoryear{Jones, Oliphant, Peterson  \& Others}{Jones
  et~al.}{2001}]{scipy}
Jones E.,  Oliphant T.,  Peterson P.,   Others 2001, {SciPy}: Open source
  scientific tools for Python, \url {http://www.scipy.org/}

\bibitem[\protect\citeauthoryear{Kass \& Raftery}{Kass \&
  Raftery}{1995}]{Kass1995}
Kass R.~E.,  Raftery A.~E.,  1995, Journal of the American Statistical
  Association, 90, 773

\bibitem[\protect\citeauthoryear{{Knutson} et~al.,}{{Knutson}
  et~al.}{2014}]{2014ApJ...785..126K}
{Knutson} H.~A.,  et~al., 2014, \mn@doi [\apj] {10.1088/0004-637X/785/2/126},
  \href {https://ui.adsabs.harvard.edu/abs/2014ApJ...785..126K} {785, 126}

\bibitem[\protect\citeauthoryear{{Lam} et~al.,}{{Lam}
  et~al.}{2017}]{2017A&A...599A...3L}
{Lam} K.~W.~F.,  et~al., 2017, \mn@doi [\aap] {10.1051/0004-6361/201629403},
  \href {https://ui.adsabs.harvard.edu/abs/2017A&A...599A...3L} {599, A3}

\bibitem[\protect\citeauthoryear{{Lehmann}, {Guenther}, {Sebastian},
  {D{\"o}llinger}, {Hartmann}  \& {Mkrtichian}}{{Lehmann}
  et~al.}{2015}]{2015A&A...578L...4L}
{Lehmann} H.,  {Guenther} E.,  {Sebastian} D.,  {D{\"o}llinger} M.,  {Hartmann}
  M.,   {Mkrtichian} D.~E.,  2015, \mn@doi [\aap]
  {10.1051/0004-6361/201526176}, \href
  {https://ui.adsabs.harvard.edu/abs/2015A&A...578L...4L} {578, L4}

\bibitem[\protect\citeauthoryear{{Lendl} et~al.,}{{Lendl}
  et~al.}{2014}]{2014A&A...568A..81L}
{Lendl} M.,  et~al., 2014, \mn@doi [\aap] {10.1051/0004-6361/201424481}, \href
  {https://ui.adsabs.harvard.edu/abs/2014A&A...568A..81L} {568, A81}

\bibitem[\protect\citeauthoryear{Line, Knutson, Deming, Wilkins  \&
  Desert}{Line et~al.}{2013}]{Line_2013}
Line M.~R.,  Knutson H.,  Deming D.,  Wilkins A.,   Desert J.-M.,  2013,
  \mn@doi [The Astrophysical Journal] {10.1088/0004-637x/778/2/183}, 778, 183

\bibitem[\protect\citeauthoryear{{Line} et~al.,}{{Line}
  et~al.}{2021}]{2021Natur.598..580L}
{Line} M.~R.,  et~al., 2021, \mn@doi [\nat] {10.1038/s41586-021-03912-6}, \href
  {https://ui.adsabs.harvard.edu/abs/2021Natur.598..580L} {598, 580}

\bibitem[\protect\citeauthoryear{{Lozovsky}, {Helled}, {Dorn}  \&
  {Venturini}}{{Lozovsky} et~al.}{2018}]{lozovsky2018}
{Lozovsky} M.,  {Helled} R.,  {Dorn} C.,   {Venturini} J.,  2018, \mn@doi
  [\apj] {10.3847/1538-4357/aadd09}, \href
  {https://ui.adsabs.harvard.edu/abs/2018ApJ...866...49L} {866, 49}

\bibitem[\protect\citeauthoryear{{Luque} et~al.,}{{Luque}
  et~al.}{2020}]{2020arXiv200711851L}
{Luque} R.,  et~al., 2020, arXiv e-prints, \href
  {https://ui.adsabs.harvard.edu/abs/2020arXiv200711851L} {p. arXiv:2007.11851}

\bibitem[\protect\citeauthoryear{Lyon \& Johnson}{Lyon \&
  Johnson}{1992}]{sesame}
Lyon S.,  Johnson J.,  1992, Technical report, Los Alamos National Laboratory,
  Los Alamos, NM.
Report LA-UR-92-3407

\bibitem[\protect\citeauthoryear{{Mancini} et~al.,}{{Mancini}
  et~al.}{2013}]{2013MNRAS.436....2M}
{Mancini} L.,  et~al., 2013, \mn@doi [\mnras] {10.1093/mnras/stt1394}, \href
  {https://ui.adsabs.harvard.edu/abs/2013MNRAS.436....2M} {436, 2}

\bibitem[\protect\citeauthoryear{{Mancini} et~al.,}{{Mancini}
  et~al.}{2014}]{2014A&A...568A.127M}
{Mancini} L.,  et~al., 2014, \mn@doi [\aap] {10.1051/0004-6361/201424106},
  \href {https://ui.adsabs.harvard.edu/abs/2014A&A...568A.127M} {568, A127}

\bibitem[\protect\citeauthoryear{{Mancini} et~al.,}{{Mancini}
  et~al.}{2019}]{2019MNRAS.485.5168M}
{Mancini} L.,  et~al., 2019, \mn@doi [\mnras] {10.1093/mnras/stz661}, \href
  {https://ui.adsabs.harvard.edu/abs/2019MNRAS.485.5168M} {485, 5168}

\bibitem[\protect\citeauthoryear{{Mankovich} \& {Fuller}}{{Mankovich} \&
  {Fuller}}{2021}]{mankovich2021}
{Mankovich} C.~R.,  {Fuller} J.,  2021, \mn@doi [Nature Astronomy]
  {10.1038/s41550-021-01448-3}, \href
  {https://ui.adsabs.harvard.edu/abs/2021NatAs...5.1103M} {5, 1103}

\bibitem[\protect\citeauthoryear{{Martins} et~al.,}{{Martins}
  et~al.}{2015}]{2015A&A...576A.134M}
{Martins} J.~H.~C.,  et~al., 2015, \mn@doi [\aap]
  {10.1051/0004-6361/201425298}, \href
  {https://ui.adsabs.harvard.edu/abs/2015A&A...576A.134M} {576, A134}

\bibitem[\protect\citeauthoryear{{Maxted} et~al.,}{{Maxted}
  et~al.}{2013}]{2013PASP..125...48M}
{Maxted} P.~F.~L.,  et~al., 2013, \mn@doi [\pasp] {10.1086/669231}, \href
  {https://ui.adsabs.harvard.edu/abs/2013PASP..125...48M} {125, 48}

\bibitem[\protect\citeauthoryear{{McCullough} et~al.,}{{McCullough}
  et~al.}{2006}]{2006ApJ...648.1228M}
{McCullough} P.~R.,  et~al., 2006, \mn@doi [\apj] {10.1086/505651}, \href
  {https://ui.adsabs.harvard.edu/abs/2006ApJ...648.1228M} {648, 1228}

\bibitem[\protect\citeauthoryear{{Melo}, {Santos}, {Pont}, {Guillot},
  {Israelian}, {Mayor}, {Queloz}  \& {Udry}}{{Melo}
  et~al.}{2006}]{2006A&A...460..251M}
{Melo} C.,  {Santos} N.~C.,  {Pont} F.,  {Guillot} T.,  {Israelian} G.,
  {Mayor} M.,  {Queloz} D.,   {Udry} S.,  2006, \mn@doi [\aap]
  {10.1051/0004-6361:20065954}, \href
  {https://ui.adsabs.harvard.edu/abs/2006A&A...460..251M} {460, 251}

\bibitem[\protect\citeauthoryear{{Miguel}, {Guillot}  \& {Fayon}}{{Miguel}
  et~al.}{2016}]{Miguel2016}
{Miguel} Y.,  {Guillot} T.,   {Fayon} L.,  2016, \mn@doi [\aap]
  {10.1051/0004-6361/201629732}, \href
  {https://ui.adsabs.harvard.edu/abs/2016A&A...596A.114M} {596, A114}

\bibitem[\protect\citeauthoryear{{Miguel} et~al.,}{{Miguel}
  et~al.}{2022}]{miguel2022}
{Miguel} Y.,  et~al., 2022, \mn@doi [\aap] {10.1051/0004-6361/202243207}, \href
  {https://ui.adsabs.harvard.edu/abs/2022A&A...662A..18M} {662, A18}

\bibitem[\protect\citeauthoryear{{Militzer} \& {Hubbard}}{{Militzer} \&
  {Hubbard}}{2013}]{mh13}
{Militzer} B.,  {Hubbard} W.~B.,  2013, \mn@doi [\apj]
  {10.1088/0004-637X/774/2/148}, \href
  {https://ui.adsabs.harvard.edu/abs/2013ApJ...774..148M} {774, 148}

\bibitem[\protect\citeauthoryear{{Mol Lous} \& {Miguel}}{{Mol Lous} \&
  {Miguel}}{2020}]{MolLous2020}
{Mol Lous} M.,  {Miguel} Y.,  2020, \mn@doi [\mnras] {10.1093/mnras/staa1405},
  \href {https://ui.adsabs.harvard.edu/abs/2020MNRAS.495.2994M} {495, 2994}

\bibitem[\protect\citeauthoryear{{M{\"u}ller}, {Helled}  \&
  {Cumming}}{{M{\"u}ller} et~al.}{2020}]{muller}
{M{\"u}ller} S.,  {Helled} R.,   {Cumming} A.,  2020, \mn@doi [\aap]
  {10.1051/0004-6361/201937376}, \href
  {https://ui.adsabs.harvard.edu/abs/2020A&A...638A.121M} {638, A121}

\bibitem[\protect\citeauthoryear{{Nettelmann} et~al.,}{{Nettelmann}
  et~al.}{2021}]{netelmann2021}
{Nettelmann} N.,  et~al., 2021, \mn@doi [The Planetary Science Journal]
  {10.3847/PSJ/ac390a}, \href
  {https://ui.adsabs.harvard.edu/abs/2021PSJ.....2..241N} {2, 241}

\bibitem[\protect\citeauthoryear{{Ni}}{{Ni}}{2020}]{Ni2020}
{Ni} D.,  2020, \mn@doi [\aap] {10.1051/0004-6361/202038267}, \href
  {https://ui.adsabs.harvard.edu/abs/2020A&A...639A..10N} {639, A10}

\bibitem[\protect\citeauthoryear{{Nikolov} et~al.,}{{Nikolov}
  et~al.}{2014}]{2014MNRAS.437...46N}
{Nikolov} N.,  et~al., 2014, \mn@doi [\mnras] {10.1093/mnras/stt1859}, \href
  {https://ui.adsabs.harvard.edu/abs/2014MNRAS.437...46N} {437, 46}

\bibitem[\protect\citeauthoryear{{Nsamba} et~al.,}{{Nsamba}
  et~al.}{2021}]{Nsamba2021}
{Nsamba} B.,  et~al., 2021, \mn@doi [\mnras] {10.1093/mnras/staa3228}, \href
  {https://ui.adsabs.harvard.edu/abs/2021MNRAS.500...54N} {500, 54}

\bibitem[\protect\citeauthoryear{{Paredes}, {Henry}, {Quinn}, {Gies},
  {Hinojosa-Go{\~n}i}, {James}, {Jao}  \& {White}}{{Paredes}
  et~al.}{2021}]{2021AJ....162..176P}
{Paredes} L.~A.,  {Henry} T.~J.,  {Quinn} S.~N.,  {Gies} D.~R.,
  {Hinojosa-Go{\~n}i} R.,  {James} H.-S.,  {Jao} W.-C.,   {White} R.~J.,  2021,
  \mn@doi [\aj] {10.3847/1538-3881/ac082a}, \href
  {https://ui.adsabs.harvard.edu/abs/2021AJ....162..176P} {162, 176}

\bibitem[\protect\citeauthoryear{{Parmentier} \& {Guillot}}{{Parmentier} \&
  {Guillot}}{2014}]{parmentier2014}
{Parmentier} V.,  {Guillot} T.,  2014, \mn@doi [\aap]
  {10.1051/0004-6361/201322342}, \href
  {https://ui.adsabs.harvard.edu/abs/2014A&A...562A.133P} {562, A133}

\bibitem[\protect\citeauthoryear{{Parmentier}, {Guillot}, {Fortney}  \&
  {Marley}}{{Parmentier} et~al.}{2015}]{parmentier2016}
{Parmentier} V.,  {Guillot} T.,  {Fortney} J.~J.,   {Marley} M.~S.,  2015,
  \mn@doi [\aap] {10.1051/0004-6361/201323127}, \href
  {https://ui.adsabs.harvard.edu/abs/2015A&A...574A..35P} {574, A35}

\bibitem[\protect\citeauthoryear{Pepper et~al.,}{Pepper
  et~al.}{2017}]{Pepper_2017}
Pepper J.,  et~al., 2017, \mn@doi [The Astronomical Journal]
  {10.3847/1538-3881/aa6572}, 153, 215

\bibitem[\protect\citeauthoryear{P\'erez \& Granger}{P\'erez \&
  Granger}{2007}]{PER-GRA:2007}
P\'erez F.,  Granger B.~E.,  2007, \mn@doi [Computing in Science and
  Engineering] {10.1109/MCSE.2007.53}, 9, 21

\bibitem[\protect\citeauthoryear{{Pinhas}, {Madhusudhan}, {Gandhi}  \&
  {MacDonald}}{{Pinhas} et~al.}{2019}]{pinhas}
{Pinhas} A.,  {Madhusudhan} N.,  {Gandhi} S.,   {MacDonald} R.,  2019, \mn@doi
  [\mnras] {10.1093/mnras/sty2544}, \href
  {https://ui.adsabs.harvard.edu/abs/2019MNRAS.482.1485P} {482, 1485}

\bibitem[\protect\citeauthoryear{{Polanski}, {Crossfield}, {Howard}, {Isaacson}
   \& {Rice}}{{Polanski} et~al.}{2022}]{jwstmetals}
{Polanski} A.~S.,  {Crossfield} I. J.~M.,  {Howard} A.~W.,  {Isaacson} H.,
  {Rice} M.,  2022, \mn@doi [Research Notes of the American Astronomical
  Society] {10.3847/2515-5172/ac8676}, \href
  {https://ui.adsabs.harvard.edu/abs/2022RNAAS...6..155P} {6, 155}

\bibitem[\protect\citeauthoryear{{Saha} \& {Sengupta}}{{Saha} \&
  {Sengupta}}{2021}]{2021AJ....162..221S}
{Saha} S.,  {Sengupta} S.,  2021, \mn@doi [\aj] {10.3847/1538-3881/ac294d},
  \href {https://ui.adsabs.harvard.edu/abs/2021AJ....162..221S} {162, 221}

\bibitem[\protect\citeauthoryear{{Sarkis}, {Mordasini}, {Henning}, {Marleau}
  \& {Molli{\`e}re}}{{Sarkis} et~al.}{2021}]{sarkis2021}
{Sarkis} P.,  {Mordasini} C.,  {Henning} T.,  {Marleau} G.~D.,   {Molli{\`e}re}
  P.,  2021, \mn@doi [\aap] {10.1051/0004-6361/202038361}, \href
  {https://ui.adsabs.harvard.edu/abs/2021A&A...645A..79S} {645, A79}

\bibitem[\protect\citeauthoryear{Sato et~al.,}{Sato et~al.}{2005}]{Sato_2005}
Sato B.,  et~al., 2005, \mn@doi [The Astrophysical Journal] {10.1086/449306},
  633, 465

\bibitem[\protect\citeauthoryear{{Sato} et~al.,}{{Sato}
  et~al.}{2012}]{2012PASJ...64...97S}
{Sato} B.,  et~al., 2012, \mn@doi [\pasj] {10.1093/pasj/64.5.97}, \href
  {https://ui.adsabs.harvard.edu/abs/2012PASJ...64...97S} {64, 97}

\bibitem[\protect\citeauthoryear{{Saumon}, {Chabrier}  \& {van Horn}}{{Saumon}
  et~al.}{1995}]{scvh}
{Saumon} D.,  {Chabrier} G.,   {van Horn} H.~M.,  1995, \mn@doi [\apjs]
  {10.1086/192204}, \href
  {https://ui.adsabs.harvard.edu/abs/1995ApJS...99..713S} {99, 713}

\bibitem[\protect\citeauthoryear{{Scaife} \& {Heald}}{{Scaife} \&
  {Heald}}{2012}]{Scaife2012}
{Scaife} A.~M.~M.,  {Heald} G.~H.,  2012, \mn@doi [\mnras]
  {10.1111/j.1745-3933.2012.01251.x}, \href
  {http://adsabs.harvard.edu/abs/2012MNRAS.423L..30S} {423, L30}

\bibitem[\protect\citeauthoryear{{Southworth}}{{Southworth}}{2010}]{2010MNRAS.408.1689S}
{Southworth} J.,  2010, \mn@doi [\mnras] {10.1111/j.1365-2966.2010.17231.x},
  \href {https://ui.adsabs.harvard.edu/abs/2010MNRAS.408.1689S} {408, 1689}

\bibitem[\protect\citeauthoryear{{Southworth}, {Tregloan-Reed}, {Pinhas},
  {Madhusudhan}, {Mancini}  \& {Smith}}{{Southworth}
  et~al.}{2018}]{2018MNRAS.481.4261S}
{Southworth} J.,  {Tregloan-Reed} J.,  {Pinhas} A.,  {Madhusudhan} N.,
  {Mancini} L.,   {Smith} A.~M.~S.,  2018, \mn@doi [\mnras]
  {10.1093/mnras/sty2488}, \href
  {https://ui.adsabs.harvard.edu/abs/2018MNRAS.481.4261S} {481, 4261}

\bibitem[\protect\citeauthoryear{{Spake} et~al.,}{{Spake}
  et~al.}{2018}]{2018Natur.557...68S}
{Spake} J.~J.,  et~al., 2018, \mn@doi [\nat] {10.1038/s41586-018-0067-5}, \href
  {https://ui.adsabs.harvard.edu/abs/2018Natur.557...68S} {557, 68}

\bibitem[\protect\citeauthoryear{{Thorngren} \& {Fortney}}{{Thorngren} \&
  {Fortney}}{2019}]{thorngren2019}
{Thorngren} D.,  {Fortney} J.~J.,  2019, \mn@doi [\apjl]
  {10.3847/2041-8213/ab1137}, \href
  {https://ui.adsabs.harvard.edu/abs/2019ApJ...874L..31T} {874, L31}

\bibitem[\protect\citeauthoryear{Thorngren, Fortney, Murray-Clay  \&
  Lopez}{Thorngren et~al.}{2016}]{thorngren2016}
Thorngren D.~P.,  Fortney J.~J.,  Murray-Clay R.~A.,   Lopez E.~D.,  2016,
  \mn@doi [The Astrophysical Journal] {10.3847/0004-637x/831/1/64}, 831, 64

\bibitem[\protect\citeauthoryear{Todorov et~al.,}{Todorov
  et~al.}{2013}]{Todorov_2013}
Todorov K.~O.,  et~al., 2013, \mn@doi [The Astrophysical Journal]
  {10.1088/0004-637x/770/2/102}, 770, 102

\bibitem[\protect\citeauthoryear{{Torres}, {Winn}  \& {Holman}}{{Torres}
  et~al.}{2008}]{2008ApJ...677.1324T}
{Torres} G.,  {Winn} J.~N.,   {Holman} M.~J.,  2008, \mn@doi [\apj]
  {10.1086/529429}, \href
  {https://ui.adsabs.harvard.edu/abs/2008ApJ...677.1324T} {677, 1324}

\bibitem[\protect\citeauthoryear{{Torres}, {Fischer}, {Sozzetti}, {Buchhave},
  {Winn}, {Holman}  \& {Carter}}{{Torres} et~al.}{2012}]{2012ApJ...757..161T}
{Torres} G.,  {Fischer} D.~A.,  {Sozzetti} A.,  {Buchhave} L.~A.,  {Winn}
  J.~N.,  {Holman} M.~J.,   {Carter} J.~A.,  2012, \mn@doi [\apj]
  {10.1088/0004-637X/757/2/161}, \href
  {https://ui.adsabs.harvard.edu/abs/2012ApJ...757..161T} {757, 161}

\bibitem[\protect\citeauthoryear{{Triaud} et~al.,}{{Triaud}
  et~al.}{2010}]{2010A&A...524A..25T}
{Triaud} A.~H.~M.~J.,  et~al., 2010, \mn@doi [\aap]
  {10.1051/0004-6361/201014525}, \href
  {https://ui.adsabs.harvard.edu/abs/2010A&A...524A..25T} {524, A25}

\bibitem[\protect\citeauthoryear{{Triaud} et~al.,}{{Triaud}
  et~al.}{2013}]{2013A&A...551A..80T}
{Triaud} A.~H.~M.~J.,  et~al., 2013, \mn@doi [\aap]
  {10.1051/0004-6361/201220900}, \href
  {https://ui.adsabs.harvard.edu/abs/2013A&A...551A..80T} {551, A80}

\bibitem[\protect\citeauthoryear{Tsiaras et~al.,}{Tsiaras
  et~al.}{2018}]{Tsiaras_2018}
Tsiaras A.,  et~al., 2018, \mn@doi [The Astronomical Journal]
  {10.3847/1538-3881/aaaf75}, 155, 156

\bibitem[\protect\citeauthoryear{{Valletta} \& {Helled}}{{Valletta} \&
  {Helled}}{2022}]{valletta2022}
{Valletta} C.,  {Helled} R.,  2022, \mn@doi [\apj] {10.3847/1538-4357/ac5f52},
  \href {https://ui.adsabs.harvard.edu/abs/2022ApJ...931...21V} {931, 21}

\bibitem[\protect\citeauthoryear{Van Der~Walt, Colbert  \& Varoquaux}{Van
  Der~Walt et~al.}{2011}]{van2011numpy}
Van Der~Walt S.,  Colbert S.~C.,   Varoquaux G.,  2011, Computing in Science \&
  Engineering, 13, 22

\bibitem[\protect\citeauthoryear{{Venturini}, {Alibert}  \& {Benz}}{{Venturini}
  et~al.}{2016}]{venturini2016}
{Venturini} J.,  {Alibert} Y.,   {Benz} W.,  2016, \mn@doi [\aap]
  {10.1051/0004-6361/201628828}, \href
  {https://ui.adsabs.harvard.edu/abs/2016A&A...596A..90V} {596, A90}

\bibitem[\protect\citeauthoryear{{Wahl}, {Hubbard}  \& {Militzer}}{{Wahl}
  et~al.}{2016}]{Wahl2016}
{Wahl} S.~M.,  {Hubbard} W.~B.,   {Militzer} B.,  2016, \mn@doi [\apj]
  {10.3847/0004-637X/831/1/14}, \href
  {https://ui.adsabs.harvard.edu/abs/2016ApJ...831...14W} {831, 14}

\bibitem[\protect\citeauthoryear{{Wahl} et~al.,}{{Wahl}
  et~al.}{2017}]{wahl2017}
{Wahl} S.~M.,  et~al., 2017, \mn@doi [\grl] {10.1002/2017GL073160}, \href
  {https://ui.adsabs.harvard.edu/abs/2017GeoRL..44.4649W} {44, 4649}

\bibitem[\protect\citeauthoryear{Wahl, Thorngren, Lu  \& Militzer}{Wahl
  et~al.}{2021}]{lovenumbers}
Wahl S.~M.,  Thorngren D.,  Lu T.,   Militzer B.,  2021, \mn@doi [The
  Astrophysical Journal] {10.3847/1538-4357/ac1a72}, 921, 105

\bibitem[\protect\citeauthoryear{Wakeford et~al.,}{Wakeford
  et~al.}{2017}]{Wakeford_2017}
Wakeford H.~R.,  et~al., 2017, \mn@doi [The Astronomical Journal]
  {10.3847/1538-3881/aa9e4e}, 155, 29

\bibitem[\protect\citeauthoryear{{Wallack}, {Knutson}  \& {Deming}}{{Wallack}
  et~al.}{2021}]{2021AJ....162...36W}
{Wallack} N.~L.,  {Knutson} H.~A.,   {Deming} D.,  2021, \mn@doi [\aj]
  {10.3847/1538-3881/abdbb2}, \href
  {https://ui.adsabs.harvard.edu/abs/2021AJ....162...36W} {162, 36}

\bibitem[\protect\citeauthoryear{{Wang} \& {Ford}}{{Wang} \&
  {Ford}}{2011}]{2011MNRAS.418.1822W}
{Wang} J.,  {Ford} E.~B.,  2011, \mn@doi [\mnras]
  {10.1111/j.1365-2966.2011.19600.x}, \href
  {https://ui.adsabs.harvard.edu/abs/2011MNRAS.418.1822W} {418, 1822}

\bibitem[\protect\citeauthoryear{{West} et~al.,}{{West}
  et~al.}{2016}]{2016A&A...585A.126W}
{West} R.~G.,  et~al., 2016, \mn@doi [\aap] {10.1051/0004-6361/201527276},
  \href {https://ui.adsabs.harvard.edu/abs/2016A&A...585A.126W} {585, A126}

\bibitem[\protect\citeauthoryear{{Wilson} et~al.,}{{Wilson}
  et~al.}{2020}]{2020MNRAS.497.5155W}
{Wilson} J.,  et~al., 2020, \mn@doi [\mnras] {10.1093/mnras/staa2307}, \href
  {https://ui.adsabs.harvard.edu/abs/2020MNRAS.497.5155W} {497, 5155}

\bibitem[\protect\citeauthoryear{Yip, Changeat, Edwards, Morvan, Chubb,
  Tsiaras, Waldmann  \& Tinetti}{Yip et~al.}{2020}]{Yip_2020}
Yip K.~H.,  Changeat Q.,  Edwards B.,  Morvan M.,  Chubb K.~L.,  Tsiaras A.,
  Waldmann I.~P.,   Tinetti G.,  2020, \mn@doi [The Astronomical Journal]
  {10.3847/1538-3881/abc179}, 161, 4

\bibitem[\protect\citeauthoryear{{von Essen}, {Mallonn}, {Welbanks},
  {Madhusudhan}, {Pinhas}, {Bouy}  \& {Weis Hansen}}{{von Essen}
  et~al.}{2019}]{2019A&A...622A..71V}
{von Essen} C.,  {Mallonn} M.,  {Welbanks} L.,  {Madhusudhan} N.,  {Pinhas} A.,
   {Bouy} H.,   {Weis Hansen} P.,  2019, \mn@doi [\aap]
  {10.1051/0004-6361/201833837}, \href
  {https://ui.adsabs.harvard.edu/abs/2019A&A...622A..71V} {622, A71}

\makeatother
\end{thebibliography}

\bsp	
\label{lastpage}

\end{document}